\documentclass[fleqn,usenatbib]{mnras}

\usepackage{newtxtext,newtxmath}
\usepackage{placeins}
\usepackage[T1]{fontenc}
\DeclareRobustCommand{\VAN}[3]{#2}
\let\VANthebibliography\thebibliography
\def\thebibliography{\DeclareRobustCommand{\VAN}[3]{##3}\VANthebibliography}

\usepackage{graphicx}	
\usepackage{amsmath}	
\usepackage{booktabs}
\usepackage{multirow}

\usepackage{xcolor}
\usepackage{xspace}

\title[NF Photo-$z$ with DESI DR1]{Beyond Point Photo-\(z\) Estimates: Normalizing Flows for Tomographic \(N(z)\) Reconstruction with DESI DR1 }

\author[Chen et al.]{
Zhuo Chen,$^{1,2}$,
Antoine Rocher$^{2}$,
Kwan Chuen Chan,$^{1,3}$\thanks{E-mail:chankc@mail.sysu.edu.cn (SYSU)},
Jean-Paul Kneib$^{2}$
\\
$^{1}$School of Physics and Astronomy, Sun Yat-Sen University, 2 Daxue Road, Tangjia, Zhuhai 519082, China\\
$^{2}$ Institute of Physics, Laboratory of Astrophysics, \'Ecole Polytechnique F\'ed\'erale de Lausanne (EPFL), Observatoire de Sauverny, CH-1290 Versoix, Switzerland \\
$^{3}$ CSST Science Center for the Guangdong-Hongkong-Macau Greater Bay Area, SYSU, Zhuhai 519082, China \\
}

\date{Accepted XXX. Received YYY; in original form ZZZ}
\pubyear{\the\year{}}
\begin{document}
\label{firstpage}
\pagerange{\pageref{firstpage}--\pageref{lastpage}}
\maketitle
\begin{abstract}

Accurate calibration of tomographic redshift distributions, $N(z)$ is essential for cosmological analysis of weak lensing data. 
We investigate conditional normalizing flows (NFs) and compare their performance with random forest (RF) and quantile random forest (QRF) models. We consider DESI galaxy and quasar targets from DESI DR1 spectroscopic data and with the photometry from Legacy Surveys DR10. We train on 1.1 million sources and test on 2.0 million sources in a non-overlapping sky region ($0<z_{\rm spec}<4.5$). For point estimation, the NF achieves the lowest scatter and outlier fraction at $0<z_{\rm spec}<1.6$ and outperforms both baselines across all three metrics at $z_{\rm spec}>1.6$. By introducing distribution-level metrics like Wasserstein distance and JS divergence, we find that stacking NF posterior PDFs provides more accurate tomographic $N(z)$ reconstruction than QRF, particularly at high redshifts. For $z_{\rm median}>2.5$, the NF achieves a smaller absolute mean-redshift bias ($0.060\pm0.031$ versus $0.318$) and higher probability overlap ($0.905\pm0.008$ versus $0.787$) than QRF. We apply redshift-dependent post-hoc PDF calibration derived from a held-out validation sample. After calibration, the absolute mean-redshift biases in four bins spanning $0.2<z_{\rm median}<1.0$ range from $0.0016$ to $0.0019$, with central values below referring benchmark  $0.002(1+z)$. In the highest-redshift bin, calibration further reduces the bias to $0.016\pm0.011$. Our results demonstrate the potential of NF PDFs for tomographic $N(z)$ reconstruction while highlighting challenges in transferring models across galaxy populations.

\end{abstract}




\section{Introduction}

Wide-field imaging surveys rely heavily on photometric redshift (photo-$z$) estimation to infer redshift information for millions to billions of galaxies.   Photo-$z$s are estimated from broadband photometry and therefore have substantially larger uncertainties than spectroscopic measurements. In particular, photo-$z$ techniques will enable the construction of enormous galaxy samples that are essential for Stage IV cosmological surveys such as Euclid ~\citep{Euclid_2011}, the Vera C. Rubin Observatory Legacy Survey of Space and Time (LSST)~\citep{lsst2019}, and the Chinese Space Station Survey Telescope (CSST) \citep{CSST_2026}. In these surveys, the accuracy of cosmological analyses depends not only on the precision of individual galaxy redshift estimates but also on the accurate reconstruction of the galaxy true redshift distribution, $N(z)$ of the sample. Weak lensing tomography and galaxy clustering analyses require precise measurement of $N(z)$ within each photometric bin. For example, the Euclid weak lensing survey requires the uncertainty in the mean redshift of each tomographic bin to be controlled at the level of $|\Delta z| \lesssim 0.002(1+z)$~\citep{EuclidNz,Euclid2021Photoz}. Achieving this level of precision represents a major challenge for photometric redshift methodologies.

In recent years, machine learning (ML) methods have become widely adopted for photometric redshift estimation~\citep{Photoz2019,Photoz2022}. ML methods can learn complex nonlinear relationships between photometric observables and galaxy redshifts, often outperforming traditional template-fitting approaches in terms of prediction accuracy and computational efficiency.  
Various algorithms have been explored including Multi-Layer Perceptron \citep{CollisterLahav_2004, Sadeh2016}, Nearest Neighbor Fitting \citep{DeVicente:2015kyp}, Random Forest \citep{Carliles_etal2010,ZhouNewman_etal2021, Lu_etal2024}, (Boosted) Decision Tree \citep{Gerdes_etal2010, CarrascoKind_Brunner2013}, Sparse Gaussian Process \citep{Almosallam_etal2016}, Convolutional Neural Network \citep{Hoyle_2016,DIsanto2018,Schuldt_etal2021,LiNapolitano_etal2022, Zhou_etal2022}, and Normalizing Flow (NF) \citep{Crenshaw_etal2024,Sun_etal2023,ren2026NF}.

Traditional ML approaches often focus on predicting a single redshift point estimate. Beyond point estimates such as the mean or median redshift, it is essential to characterize the uncertainty in redshift predictions, which can be represented by the full redshift probability density function (PDF). Normalizing Flows (NFs) \citep{Tabak:2013cnz,RezendeMohamed2015} provide a flexible framework for modelling complex probability distributions through a sequence of invertible transformations from a simple base distribution to a target distribution. Their ability to perform exact likelihood evaluation, efficient sampling, and expressive modelling of non-Gaussian and multimodal densities makes them well suited to photo-$z$ estimation, where redshift PDFs can exhibit asymmetric shapes, extended tails, and multiple peaks due to photometric degeneracies.

Accurate reconstruction and calibration of tomographic
redshift distributions, $N(z)$, are essential for reliable
cosmological inference, as systematic errors in these
distributions can propagate into cosmological parameter
estimates. Existing approaches to redshift-distribution
calibration include self-organizing maps (SOM)
\citep{KindBrunner_2014,Masters_etal2015,Buchs_etal2019,
Wright_etal2020} and clustering-based methods.
The latter can be divided into clustering-redshift
approaches using external spectroscopic reference
samples \citep{Newman_2008,McQuinnWhite_2013,
Menard_etal2013,Schmidt_etal2013}, self-calibration
methods using photometric clustering alone
\citep{Schneider_etal2006,Zhang_etal2010,
Zhang_etal2017,Peng_etal2022,XU2023}, and
combinations of these approaches \citep{Zheng_etal2024}.
Clustering-redshift techniques have been widely applied
in different surveys
\citep{Gatti_etal2018,Cawthon_etal2022,
Hildebrandt_etal2021,Rau_etal2023,ChoppindeJanvry2026}.
However, recovering the detailed shape of $N(z)$ can
be challenging due to degeneracies with galaxy bias
and other observational effects. In this context,
redshift distributions reconstructed from photometric
redshift PDFs can provide complementary information
that may be combined with clustering-based constraints.

Stacking individual photo-$z$ PDFs has been widely
used to reconstruct redshift distributions,
particularly for weak-lensing tomography
\citep{Bordoloi2010,Bonnett2015,Euclid2021Photoz}.
More recently, \citet{ren2026NF} applied conditional
NFs to photo-$z$ estimation, focusing primarily on
point-estimate accuracy, while \citet{2026tian}
developed a discretized photo-$z$ PDF estimator
using spectroscopic samples including DESI DR1
and evaluated the overall stacked $N(z)$.
Building on these studies, we systematically
quantify the performance of continuous NF PDFs
in tomographic $N(z)$ reconstruction across
photometric-redshift bins, compare their
performance with QRF, and assess the impact
of post-hoc PDF calibration, including at high
redshifts.

To investigate the potential of NFs for applications
in large galaxy surveys, we construct million-scale
datasets by combining spectroscopic redshifts from
DESI Data Release 1 (DR1) \citep{DESI2026DR1}
with photometric measurements from the DESI
Legacy Surveys \citep{Desi2019}.
We evaluate the models using independent
spectroscopic testing samples from non-overlapping
sky regions, allowing us to assess probabilistic
photo-$z$ estimation and $N(z)$ reconstruction
within a controlled spectroscopic sample.
We also compare the NF with Random Forest (RF)
and Quantile Random Forest (QRF) baselines.

Importantly, the NF framework is not inherently
specific to DESI, but provides a flexible methodology
for probabilistic photo-$z$ estimation that can be
adapted to different photometric surveys.
The conditioning features can be adjusted according
to the available photometric measurements, allowing
the framework to be retrained and calibrated for
different galaxy populations and observational
datasets, including those from Euclid and LSST.
However, the performance obtained from the
DESI-selected spectroscopic samples should not be
interpreted as the expected accuracy for the broader
photometric populations of these surveys.
Extending the methodology to such populations
will require careful assessment of spectroscopic
representativeness and selection effects.

This paper is organized as follows: Section~\ref{sec:data} describes the DESI DR1 spectroscopic sample and the Legacy Surveys DR10 photometric data, while Section~\ref{sec:method} introduces the NF, RF, and QRF methods and the $N(z)$ reconstruction metrics. Section~\ref{sec:results} presents the main results, including point-estimation performance, tomographic reconstruction, PDF calibration, and high-redshift tests. We discuss the implications and limitations of the NF approach in Section~\ref{sec:discuss} and summarize our conclusions in Section~\ref{sec:conclusion}.
\section{Datasets}
\label{sec:data}
\subsection{Spectroscopic data}
In this work, we rely on the spectroscopic measurement from the Dark Energy Spectroscopic Instrument (DESI;~\cite{DESI2022}), an eight-years program that aims to observe over 60 million galaxies and quasars across approximately $17\,000\,\mathrm{deg}^2$ of the extragalactic sky. Its first public Data Release 1 (DR1; \cite{DESI2026DR1}) contains approximately 18.7 million high-quality spectra from the Main Survey, including about 13.0 million galaxies and 1.6 million quasars with well-characterized uncertainties and quality flags. The large sample size and broad redshift coverage of DESI DR1 provide a valuable dataset for training and evaluating probabilistic photo-$z$ models, with spectroscopic redshift distributions, $N_{\rm spec}(z)$, serving as references for assessing the accuracy of reconstructed $N(z)$.

\subsection{Photometric features}
\label{subsec:photometric feature}
The DESI target selection is primarily based on colour
and magnitude information from the DESI Legacy Imaging
Surveys DR9 \citep{DESI2022}, which combine optical
imaging from the Dark Energy Camera Legacy Survey
(DECaLS), the Beijing--Arizona Sky Survey (BASS),
and the Mayall z-band Legacy Survey (MzLS) with
infrared observations from WISE
\citep{Wright2010WISE,Desi2019}.

For our analysis, we use the Legacy Surveys DR10.1
release\footnote{At the beginning of this analysis,
DR11 was not yet available.}, which additionally
includes DECam $i$-band imaging
\citep{Flaugher2015DECam}, providing extended
wavelength coverage for photometric redshift estimation.

The DESI spectroscopic sample comprises several
target classes with different selection criteria.
The main magnitude limits include
$r<19.5$ for BGS Bright,
$19.5<r<20.175$ for BGS Faint,
$z_{\rm fiber}<21.6$ for LRGs,
$g_{\rm fiber}<24.1$ for ELGs,
and $16.5<r<23.0$ for QSOs
\citep{hahn_desi_2023,zhou_target_2023,
raichoor_target_2023,chaussidon_target_2022}.
These magnitude limits, together with the additional
colour and quality selections, imply that the DESI
spectroscopic sample is not necessarily representative
of the broader photometric galaxy population, as it
preferentially samples specific regions of
colour--magnitude space.

As input features to train the photometric redshift calibration, we use fluxes and inverse variances (IVAR) in the $g$, $r$, $i$, $z$, $W1$, and $W2$ bands. The fluxes are converted into extinction-corrected magnitudes as
\begin{equation}
m_{\mathrm{band}} = 22.5 - 2.5\log_{10}\left(\frac{F_{\mathrm{band}}}{T_{\mathrm{MW,band}}}\right),
\end{equation}
where $T_{\mathrm{MW,band}}$ accounts for Galactic extinction using the extinction coefficients from~\citet{Schlegel1998Dust,Schlafly2011extinction}. Additionally, we include the colours $g-r$, $r-i$, $i-z$, $z-W1$ and $W1-W2$, which provide complementary information. Finally we incorporate galaxy properties beyond photometric information using morphological and observational features including the half-light radius (\texttt{shape\_r}), S\'ersic index (\texttt{sersic}), and PSF depths in the $g$, $r$, $i$, and $z$ bands.

\subsection{Training and testing datasets}
\label{subsec:data}
For this analysis, we construct an initial parent sample of approximately $3\times10^7$ sources from the DESI Legacy Imaging Surveys DR10 photometric catalogue~\citep{Desi2019}. We identify sources with corresponding spectroscopic information from DESI DR1 to construct the dataset for photometric-redshift estimation. We then apply the following quality selections to ensure reliable photometric and spectroscopic measurements:
\begin{enumerate}
    \item sources are DESI galaxy or QSO targets
    
    \item Positive flux measurements in all bands:
    \[
    \mathrm{Flux}_{\rm all}>0 .
    \]
    \item Reliable spectroscopic redshifts:
    \[
    \mathrm{ZCAT\_PRIMARY}=\mathrm{True},\quad \mathrm{ZWARN}=0 .
    \]
    \item Signal-to-noise ratio greater than 3 in all bands:
    \[
    \mathrm{SNR}_{\rm Flux_{all}}>3 .
    \]
    
\end{enumerate}

After these selections, about $6\times10^6$ objects remain, including stars, galaxies and quasars with redshift ranging around $0<z<4.5$. Instead of training separate models for different galaxy populations or source classes (e.g. Bright Galaxies, Luminous Red Galaxies, and quasars), we train a single model on the full sample. This allows the NF to learn a unified mapping between photometric observables and redshift without introducing additional selection effects associated with population-specific pre-selection. We also do not remove stars or other contaminants using spectroscopic information, such as \(z_{\rm spec}\) or spectroscopic classifications. This choice is intended to ensure that the pipeline can be applied directly to realistic imaging surveys, where such spectroscopic information is not available for pre-selection. 

\begin{figure*}
    \centering
    \includegraphics[width=0.49\linewidth]{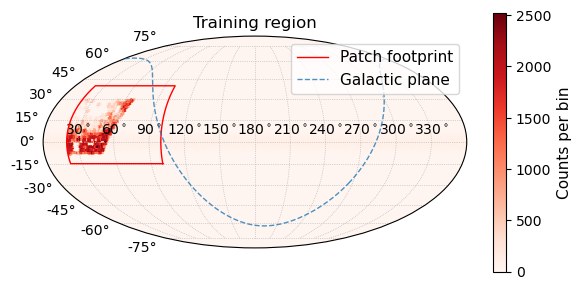}
   \includegraphics[width=0.49\linewidth]{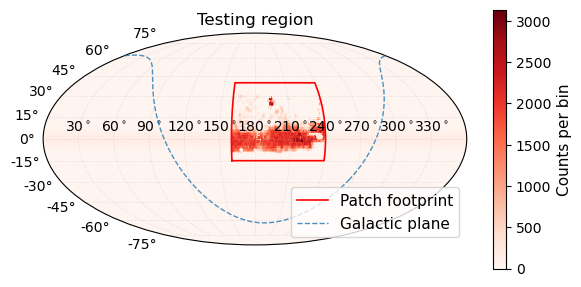}
    \caption{Spatial footprints and number densities of the training and testing patches in equatorial coordinates using a Mollweide projection. The red lines indicate the adopted patch boundaries, while the blue dashed line shows the Galactic plane.}
    \label{fig:footprint}
\end{figure*}

To evaluate the transferability of the model across different sky regions, we divide the dataset into two spatially separated areas within the DESI footprint. The first region (\(80^\circ<\mathrm{RA}<160^\circ\), \(-15^\circ<\mathrm{DEC}<40^\circ\)) is used for training and validation, while the second region (\(160^\circ<\mathrm{RA}<240^\circ\), \(-15^\circ<\mathrm{DEC}<40^\circ\)) is reserved as an independent test set, as shown in Fig.~\ref{fig:footprint}. The normalised distributions of photometric features and observational features for these sources in two patches are shown in the Fig.~\ref{fig:featuredistributions} in Appendix.

\begin{figure}
    \centering
    \includegraphics[width=\linewidth]{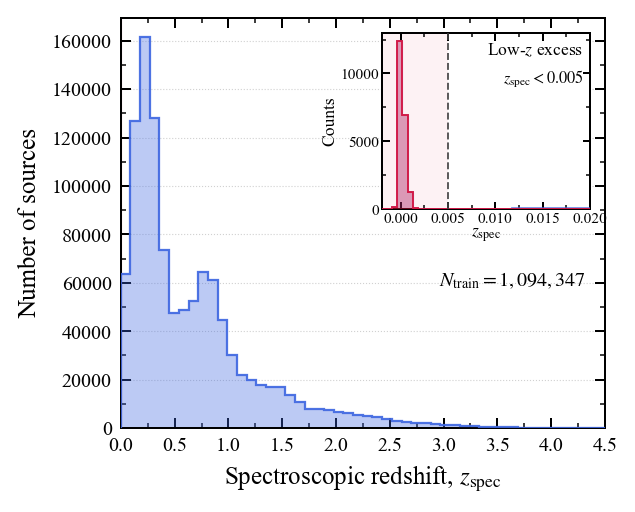}
    \includegraphics[width=\linewidth]{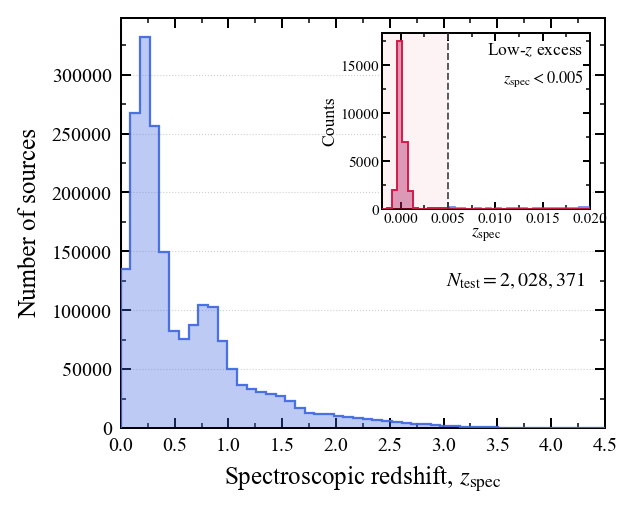}
    \caption{Upper panel: Spectroscopic redshift distribution of the training sample
  ($N_{\rm train}=1{,}094{,}347$). Lower panel: Spectroscopic redshift distribution of the testing sample $N_{\rm test}=2{,}028{,}371$. The inset panels enlarge the very low-redshift range, $-0.002<z_{\rm spec}<0.02$, highlighting the pronounced excess at $z_{\rm spec}<0.005$, which is primarily due to stellar contamination. }
    \label{fig:Nztraining}
\end{figure}

Within the training region, 10\% of the objects are randomly selected to form a validation set, while the remaining 90\% are used for training. The final dataset comprises $1{,}094{,}347$ training sources, $121{,}594$ validation sources, and the testing set has $2{,}028{,}371$ sources (including 252,719 quasars and 25,511 stars). The spectroscopic redshift distributions of the training and testing samples are shown in Fig. ~\ref{fig:Nztraining}. Both samples exhibit a pronounced excess at very low redshift ($z_{\rm spec}<0.005$), which is mainly due to stellar contamination. 

\section{Methods}
\label{sec:method}

In this section, we first describe the three methods used for photo-$z$ estimation: NF, random forest and quantile random forest. We then introduce the photometric redshift distributions by stacking PDFs and the metrics used to quantify its accuracy relative to the ground spectroscopic redshift distribution.

\subsection{Normalizing flows}

Below we briefly review the method used for photo-\(z\) estimation and refer the reader to~\citet{ren2026NF} for more details. NFs are generative probabilistic models that construct complex probability distributions by applying a sequence of invertible and differentiable transformations to a simple base distribution~\citep{Dinh2014NICE}. Their tractable change-of-variables formulation allows both exact likelihood evaluation and efficient sampling.

For photo-$z$ estimation, we use a conditional NF to model the redshift distribution
\(
p(z\mid\mathbf{x}_{\rm phot}),
\)
where $\mathbf{x}_{\rm phot}$ denotes the input photometric features.
Let $u$ denote a latent variable drawn from a simple base distribution,
$p_U(u)$, which is taken to be Gaussian. The conditional flow defines an
invertible transformation
\(
z = g(u;\mathbf{x}_{\rm phot} )
\)
with inverse
\(
u = f(z;\mathbf{x}_{\rm phot} ).
\)
The conditional density is then obtained from the change-of-variables
formula,
\begin{equation}
p(z\mid\mathbf{x}_{\rm phot})
=
p_U\ \left( f(z;\mathbf{x}_{\rm phot} ) \right)
\left|
\frac{d f}
{d z}
\right|.
\end{equation}
The flow is composed of a sequence of invertible transformations, and the
model parameters are optimized by maximizing the mean conditional
log-likelihood of the training sample,
\begin{equation}
\mathcal{L}
=
\frac{1}{N}
\sum_{j=1}^{N}
\log p\!\left(
z_j\mid\mathbf{x}_{{\rm phot},j} \right),
\label{eq:lossfunction}
\end{equation}
where $N$ is the number of training targets.

We adopt the same network implementation and hyperparameters as~\cite{ren2026NF}. Following \citet{ren2026NF}, we use the Neural Spline Flows (NSFs)~\citep{NF2019} to implement the flow transformation for its capability to express complex transformation and hence superior performance in redshift estimation.  
In practice, the NF network uses a standard normal base distribution and three conditional neural spline layers, each parameterized by an MLP with three hidden layers of 32 neurons and a spline bound of $6.0$. The model is optimized using the Adam optimizer with a weight decay of \(10^{-4}\), with gradient clipping at 10.

Before training, the spectroscopic redshifts and input photometric features are standardised using z-score normalisation to have zero mean and unit variance. The model is trained for 75 epochs, with the learning rate reduced by a factor of 10 after 50 epochs. Training is performed by maximising the mean log-likelihood defined in Eq.~\ref{eq:lossfunction}, or equivalently, minimising its negative value. The validation loss is monitored throughout training to assess potential overfitting. Towards the end of training, both the training and validation losses fluctuated around approximately stable values without sustained improvement and remained close to each other, motivating the adopted training length of 75 epochs.

Training the model for 75 epochs with approximately $1.1$ million sources requires about 6 hours on an NVIDIA RTX 4060 GPU. To assess training consistency, we evaluate the final loss values of five independently trained NF models on the same training and validation samples. In this case, the resulting mean training and validation losses are $-1.1499 \pm 0.0234$ and $-1.1488 \pm 0.0238$, respectively, where the uncertainties represent the standard deviations across the five realisations with different random seeds. The close agreement between the training and validation losses suggests no substantial generalisation gap in the final models.

For a galaxy with photometric features $\mathbf{x}_{\rm phot}$, the trained model predicts the conditional redshift PDF,
\(
p(z\mid\mathbf{x}_{\rm phot}).
\)
The full predicted PDF can be retained for probabilistic analyses or summarised using point estimates. We adopt the posterior median, $z_{\rm median}$, as the primary point estimate because of its robustness to asymmetric and multimodal distributions. For each object in the test sample, we draw $N_{\rm sample}=200$ Monte Carlo samples from its predicted conditional distribution:
\begin{equation}
z \sim p(z\mid\mathbf{x}_{\rm phot}).
\label{eq:nf_mc_samples}
\end{equation}
We adopt fewer samples than the $N_{\rm sample}=2000$ used by \citet{ren2026NF} to reduce the computational cost for the large testing dataset.

\subsection{Random Forest and Quantile Random Forest}

Additionally, we use Random Forest (RF;~\cite{RF2001}) and Quantile Random Forest (QRF;~\cite{QRF2006}) as baseline methods because of their computational efficiency, robustness, and widespread use in astronomical applications~\citep{Photoz2019,Photoz2022}. RF is an ensemble-learning method that combines the predictions of multiple decision trees. For photo-$z$ estimation, each tree is trained using photometric features and spectroscopic redshifts, and the final point estimate is obtained by averaging the predictions over the ensemble. QRF extends RF by estimating conditional quantiles of the target variable rather than only its conditional mean. These quantiles provide a discrete representation of the conditional redshift distribution. Increasing the number of quantiles can improve the resolution of this representation, although estimates at extreme or closely spaced quantile levels may become noisy in sparsely sampled regions of the photometric feature space. Consequently, QRF generally provides less smooth PDFs than those produced by the NF model.

The RF is implemented using the \texttt{RandomForestRegressor} class from \texttt{scikit-learn}\footnote{\url{https://scikit-learn.org/stable/}}. To train the RF, we use 500 decision trees with \texttt{max\_depth=None}, \texttt{min\_samples\_leaf=5}, and \texttt{max\_features="sqrt"}. Missing input values are handled using median imputation. The QRF is implemented using the \texttt{quantile\_forest} package\footnote{\url{https://zillow.github.io/quantile-forest/}}. The QRF uses 500 decision trees with \texttt{max\_depth=20}, \texttt{min\_samples\_leaf=5}, and \texttt{max\_features="sqrt"}. Its conditional redshift distribution is evaluated at 99 uniformly spaced quantile levels between 0.01 and 0.99.  Both models use the same training and testing samples as the NF model (see Sec.~\ref{subsec:data}), ensuring a consistent comparison.

\subsection{Redshift-distribution reconstruction and evaluation metrics}
\label{subsec:nz_reconstruction}

The photometric redshift distribution is reconstructed by stacking the normalised PDFs of individual source ~\citep{Bordoloi2010,Bonnett2015}:
\begin{equation}
N_{\rm phot}(z)
=
\sum_{i=1}^{N_{\rm gal}} P_i(z),
\label{eq:stackingPDFs}
\end{equation}
where $P_i(z)$ is the normalised redshift PDF of the $i^\mathrm{th}$ source. For methods that provide only point estimates, such as RF, the distribution is replaced by the Dirac delta function distribution.  
In practice, the point estimates are binned using the same redshift bins as the PDF-based reconstruction.

To evaluate the performance metrics of the photo-$z$ point estimation, we use the normalised redshift
error
\begin{equation}
\Delta z =
\frac{z_{\rm phot}-z_{\rm spec}}
{1+z_{\rm spec}},
\end{equation}
and three commonly used summary statistics~\citep{Photoz2019}. The
scatter is quantified by the normalised median absolute deviation,
\begin{equation}
\sigma_{\rm NMAD}
=
1.4868\,{\rm median}
\left(
\left|
\Delta z-{\rm median}(\Delta z)
\right|
\right),
\end{equation}
while the absolute median bias is defined as
\begin{equation}
\left|{\rm bias}\right|
=
\left|
\rm{median}(\Delta z)
\right|.
\end{equation}
We further define catastrophic outliers as sources satisfying
\begin{equation}
|\Delta z|>0.15.
\end{equation}

The point estimates of the NF model are obtained from the predicted posterior $p(z\mid\mathbf{x})$ using its mean, median, or mode ($z_{\rm mean}$, $z_{\rm median}$, and $z_{\rm maxp}$, respectively)~\citep{Crenshaw_etal2024, ren2026NF}. We adopt $z_{\rm median}$ as the primary estimator because of its robustness to asymmetric and multimodal PDFs as \citet{ren2026NF}.

Furthermore, we introduce several performance metrics to compare the reconstructed photometric redshift distribution, $N_{\rm phot}(z)$ and the spectroscopic reference, $N_{\rm spec}(z)$. Both distributions are normalised to unit integral before these metrics are evaluated.
The mean-redshift bias is defined as
\begin{equation}
\Delta z_{\rm mean}
=
\
\langle z\rangle_{\rm phot}
-
\langle z\rangle_{\rm spec}
,
\end{equation}
where $\langle z\rangle_{\rm phot}$ and $\langle z\rangle_{\rm spec}$ are respectively the mean redshifts of photometric redshift distribution $N_{\rm phot}(z)$ and spectroscopic redshift distribution $N_{\rm spec}(z)$. This quantifies the displacement of the reconstructed distribution relative to the spectroscopic reference. We also measure the difference of their widths using
\begin{equation}
\Delta\sigma
=
\
\sigma_{\rm phot}
-
\sigma_{\rm spec}
,
\end{equation}
where $\sigma_{\rm phot}$ and $\sigma_{\rm spec}$ are the standard deviation of both distributions.
Differences in the overall distribution shape are quantified using the first-order Wasserstein distance~\citep{Villani2009wat1}:
\begin{equation}
W_1
=
\int
\left|
F_{\rm phot}(z)-F_{\rm spec}(z)
\right|
\,{\rm d}z,
\end{equation}
where $F_{\rm phot}$ ($F_{\rm spec}$) denotes the Cumulative Distribution Function (CDF) derived from $N_{\rm phot}$ ($N_{\rm spec}$). This metric is sensitive to shifts and broad distortions of the reconstructed distribution.
Additionally, we use the Jensen--Shannon (JS) divergence~\citep{Lin1991JW},
\begin{equation}
{\rm JS}(  N_{\rm phot} \parallel N_{\rm spec} )
=
\frac{1}{2}D_{\rm KL}( N_{\rm phot} \parallel M)
+
\frac{1}{2}D_{\rm KL}( N_{\rm spec} \parallel M),
\end{equation}
where $M=\frac{1}{2}(N_{\rm phot} + N_{\rm spec})$ and $D_{\rm KL} $ is the Kullback-Leibler divergence \citep{Kullback1951overlap}. The JS divergence is symmetric and measures differences in both the central and tail regions of the distributions. Finally, the shared probability mass is quantified using the overlap coefficient
\begin{equation}
{\rm Overlap}
=
\int
\min\left[
N_{\rm phot}(z),
N_{\rm spec}(z)
\right]
\,{\rm d}z.
\end{equation}
For distributions normalised to unit integral, the overlap ranges from 0 to 1, with larger values indicating closer agreement. Together, these metrics characterize differences in the location, width, shape, and overall agreement between the reconstructed redshift distribution, $N_{\rm phot}(z)$, and the spectroscopic reference distribution, $N_{\rm spec}(z)$.

\section{Results}
\label{sec:results}

\subsection{Point-estimation performance}

The 3 models are trained with $\sim1.1$ million sources, using the photometric magnitudes and colours ($g,r,i,z,W1,W2,g-r,r-i,i-z,z-W1,W1-W2$), together with the morphological and observational features described in Section~\ref{subsec:photometric feature}. The influence of the individual photometric features for NF model is presented in Appendix~\ref{app:features}, where the colours like $g-r$, $r-i$ and $i-z$ show strongest sensitivity in NF model training based on Fig.~\ref{fig:saliency}. We use \(z_{\rm median}\) as the point estimate for both the NF and QRF models, while the RF directly provides a single redshift prediction. 

The quantitative comparison of point-estimation performance is summarized in Table~\ref{tab:pointestimation} for two redshift ranges, \(0<z_{\rm spec}<1.6\) dominated by galaxies, and \(z_{\rm spec}>1.6\) dominated by quasars. At low redshift, the NF achieves the lowest scatter, \(\sigma_{\rm NMAD}=0.0223\pm0.0002\), and outlier fraction, \((3.79\pm0.03)\%\), while maintaining a competitive absolute median bias of \((6.54\pm5.40)\times10^{-4}\). The QRF yields the smallest median bias, \(4.8\times10^{-5}\). At high redshift, the NF performs best across all three metrics, with an absolute median bias of \(0.0430\pm0.0034\), \(\sigma_{\rm NMAD}=0.1015\pm0.0011\), and an outlier fraction of \((25.74\pm0.51)\%\), indicating a clearer advantage in the more challenging high-redshift regime. The comparison between photometric and spectroscopic redshifts for the three methods is shown in Fig.~\ref{fig:QRF and RF match} in the Appendix. The reported uncertainties for NF are estimated as the standard deviations across five independent training realisations with different random seeds and data shuffling. The uncertainties for RF and QRF are not reported because the corresponding seed-to-seed variations are below the displayed precision. To summarise, the NF is competitive for point estimation, with a clearer advantage at high redshift, but it does not consistently outperform RF or QRF in the low-redshift regime.   Our results agree with the findings in \citet{ren2026NF}, which applied the conditional NF  on different datasets and concluded that NF excels when the data are noisy and sparse by comparing with various photo-$z$ estimation methods.  

The main strength of NF becomes more apparent when the full predicted PDFs are retained, as they preserve information on the uncertainty, width, asymmetry, and extended tails of the redshift distribution.

\begin{table}
    \centering
    \caption{Summary of the point-estimation performance metrics for the Normalizing Flow (NF), quantile random forest (QRF), and random forest (RF) models. For NF and QRF, the median of the predicted redshift distribution, \(z_{\rm median}\), is adopted as the point estimate, while for RF we use the default ensemble-averaged prediction. We report the metrics in two redshift ranges: \(0<z_{\rm spec}<1.6\), containing 1,880,915 sources, and \(z_{\rm spec}>1.6\), containing 135,522 sources. The uncertainties for NF are estimated as the standard deviations across five independent training realisations with different random seeds and data shuffling. Some uncertainties are not reported because the corresponding seed-to-seed variations are below the displayed precision.}
\setlength{\tabcolsep}{4pt}
\begin{tabular}{@{}lcccc@{}}
\hline
Redshift & Model &
$|\mathrm{bias}|$ ($10^{-4}$) &
$\sigma_{\rm NMAD}$ ($10^{-2}$) &
$f_{\rm out}$ (\%) \\
\hline

\multirow{3}{*}{$0<z_{\rm spec}<1.6$}
& NF  & $6.54\pm5.40$ & $2.23\pm0.02$ & $3.79\pm0.03$ \\
& RF  & $16.47$ & $2.72$ & $4.78$ \\
& QRF & $0.48$ & $2.69$ & $3.99$ \\

\hline

\multirow{3}{*}{$z_{\rm spec}>1.6$}
& NF  & $430\pm34$ & $10.15\pm0.11$ & $25.74\pm0.51$ \\
& RF  & $794$ & $11.89$ & $31.49$ \\
& QRF & $631$ & $11.22$ & $28.88$ \\

\hline
\end{tabular}
\label{tab:pointestimation}
\end{table}

.

\subsection{Tomographic N(z) reconstruction}
\label{sec:tomoN(z)}
In this subsection, we present the reconstruction results for the overall and tomographic redshift distributions, $N(z)$, obtained by stacking the predicted photo-$z$ PDFs. We compare the performance of the NF and QRF models using the metrics defined in Section~\ref{subsec:nz_reconstruction}, with the spectroscopic redshift distributions from DESI DR1 as the reference. 

\begin{figure}
    \centering
\includegraphics[width=0.95\linewidth]{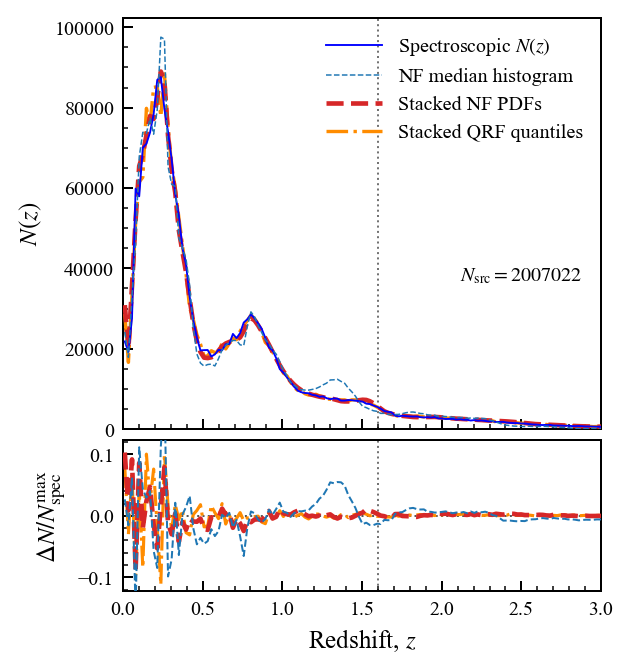}
   \caption{
Comparison of the reconstructed redshift distributions,
$N(z)$, for the testing sample
displayed over $0<z<3$ for clarity. The upper panel
shows the spectroscopic reference distribution alongside
the reconstructions obtained by stacking the normalizing-flow
(NF) and quantile random forest (QRF) PDFs, as well as
the distribution of NF posterior-median redshift estimates.
The lower panel shows the corresponding residuals
relative to the spectroscopic distribution, normalised
by its maximum value, $\Delta N/N_{\rm spec}^{\rm max}$,
where $\Delta N=N_{\rm pred}-N_{\rm spec}$.
The vertical dotted line at $z=1.6$ approximately
separates the galaxy-dominated and quasar-dominated
redshift ranges.
}
 \label{fig:Nzreconstruction}
\end{figure}

Firstly, we consider reconstructing the overall redshift distributions of testing sources, $N(z)$, by stacking the predicted photo-$z$ PDFs over the redshift range $0<z<4.5$, using 200 uniformly spaced bins. We compare the NF and QRF reconstructions with the reference redshift distribution derived from DESI DR1 spectroscopic redshifts. We also display the redshift distribution constructed from the NF posterior-median estimates to illustrate the differences between PDF stacking and point-estimate reconstruction. We do not include the RF results in the $N(z)$ comparison, as RF provides only point estimates and does not support PDF stacking. 200 Monte Carlo samples are used to estimate the PDF for each sources with the NF models, whereas the QRF distribution is represented by 99 quantiles uniformly spaced from 0.01 to 0.99. For NF, we checked the variance of the 200 Monte Carlo samples and it was negligible compare to the run-to-run case.
For comparison, we construct $N(z_{\rm median})$ using the posterior median of each galaxy, shown in Fig.~\ref{fig:Nzreconstruction}.

\begin{figure*}
    \centering
    \includegraphics[width=0.80\linewidth]{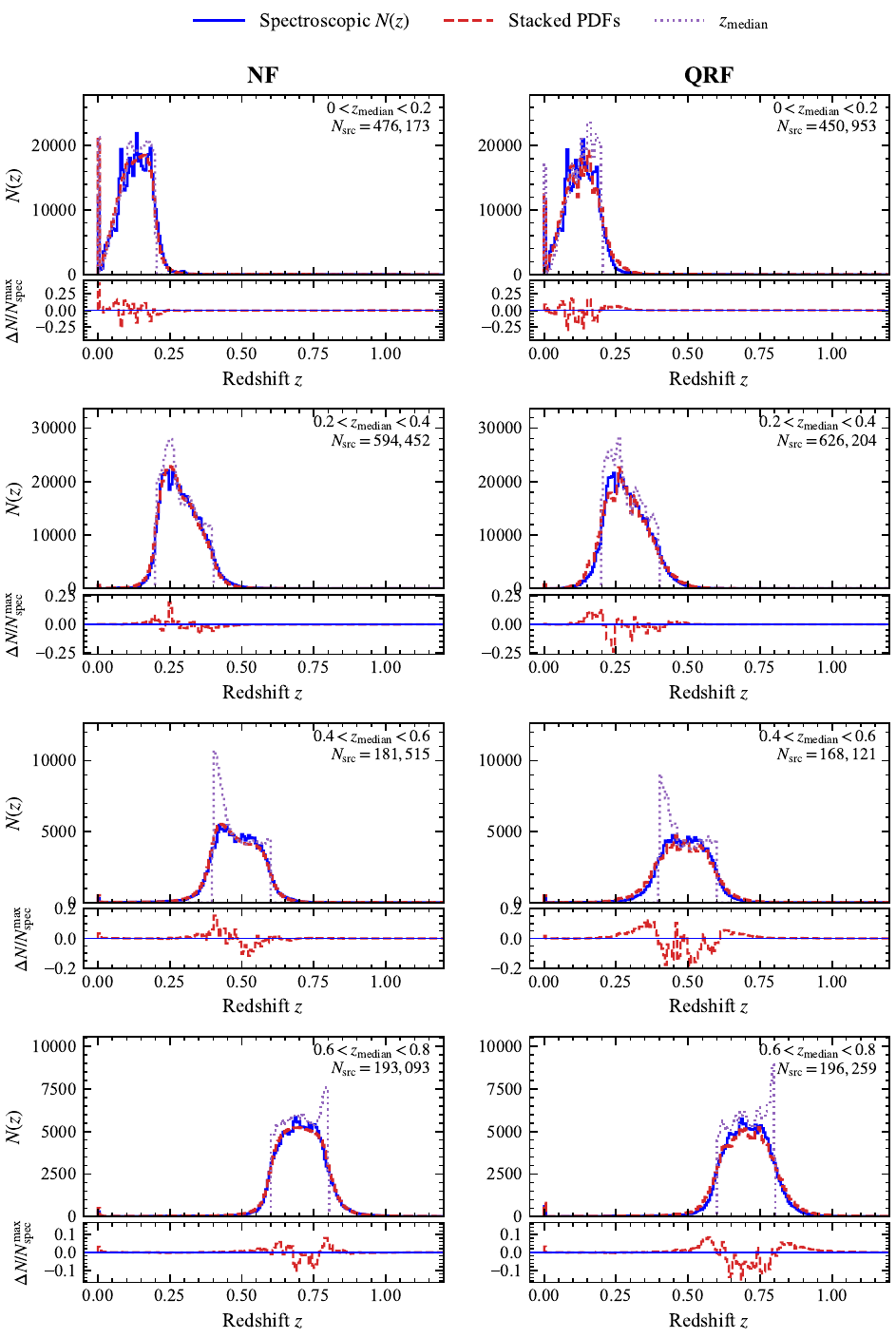}
\caption{
Tomographic redshift distributions, $N(z)$, reconstructed by stacking the predicted redshift PDFs from the NF (left) and QRF (right). The samples are divided into eight tomographic bins according to the median of the corresponding predicted redshift distribution, $z_{\rm median}$: $0.0<z_{\rm median}<0.2$ (top), $0.2<z_{\rm median}<0.4$ (second row), $0.4<z_{\rm median}<0.6$ (third row), and $0.6<z_{\rm median}<0.8$ (bottom). For all four tomographic bins, $N(z)$ is evaluated over $0<z<1.2$ using 200 uniform redshift bins. The dotted histograms show the distributions of the point estimates, $z_{\rm median}$, the solid blue lines show the spectroscopic reference distributions, $N_{\rm spec}(z)$, and the dashed red lines show the reconstructed distributions obtained by stacking the predicted PDFs, $N_{\rm pred}(z)$. The lower sub-panels show the normalised residuals, $\Delta N/N_{\rm spec}^{\rm max}$, where $\Delta N=N_{\rm pred}-N_{\rm spec}$.
}
    \label{fig:Nzthreebins}
\end{figure*}

\begin{figure*}
  \includegraphics[width=0.80\linewidth]{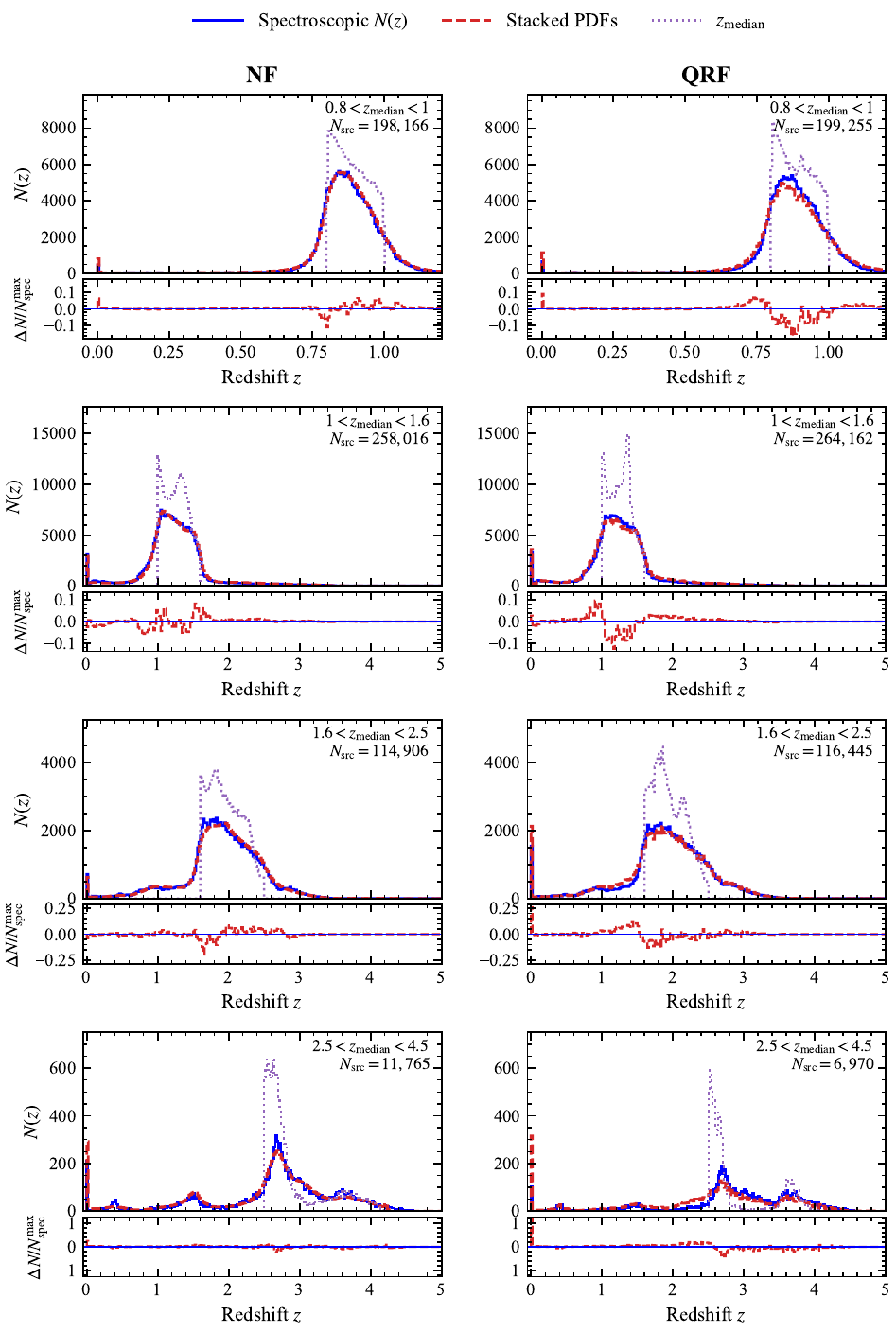}
       \caption{Same as Fig.~\ref{fig:Nzthreebins} for the four bins defined by $z_{\rm median}$: $0.8<z_{\rm median}<1.0$, $1.0<z_{\rm median}<1.6$, $1.6<z_{\rm median}<2.5$, and $z_{\rm median}>2.5$ from top to bottom. The first row uses $0<z<1.2$ with 200 uniform bins, while the remaining rows use $0<z<5.0$ with 250 bins. Dotted histograms show the $z_{\rm median}$ distributions, solid blue lines the spectroscopic $N_{\rm spec}(z)$, and dashed red lines the stacked-PDF reconstruction, $N_{\rm pred}(z)$. Lower panels show $\Delta N/N_{\rm spec}^{\rm max}$, where $\Delta N=N_{\rm pred}-N_{\rm spec}$.
}
\label{fig:Nzhighbins}
\end{figure*}

For tomographic analysis, we compare the tomographic $N(z)$ reconstructions obtained by stacking the NF and QRF redshift distributions. The test samples are divided according to the predicted posterior-median redshift with boundaries [0.0, 0.2, 0.4, 0.6, 0.8, 1.0, 1.6, 2.5, 4.5] in 8 tomographic bins. For the first 5 bins, the tomographic redshift distributions $N(z)$ are reconstructed on the redshift range $0.0<z<1.2$ with 200 uniform bins. For the remaining 3 bins, the tomographic redshift distributions $N(z)$ are reconstructed on the redshift range $0.0<z<5.0$ with 250 uniform bins. In practice, the tomographic bins are defined using the NF $z_{\rm median}$ estimates for NF model, while the QRF reconstruction uses the corresponding QRF median estimates. This provides an end-to-end comparison of the two pipelines, in which each method defines its own tomographic sample using its predicted median redshift, more closely reflecting a realistic photo-\(z\) tomographic analysis. The resulting NF and QRF reconstructions are shown in Fig.~\ref{fig:Nzthreebins} and Fig.~\ref{fig:Nzhighbins}. 

The quantitative comparison is summarized in Table~\ref{tab:Nzmetrics_bins}. Each NF metric is averaged over five independent training realisations using the same training and testing samples but different random shuffling and seeds. The quoted uncertainty is the standard deviation across these realisations. For the QRF, the seed-to-seed variations are negligible compared with the NF run-to-run uncertainties and are therefore not reported.


 \begin{table*}
\centering
\caption{
Comparison of the tomographic and full-sample \(N(z)\) reconstructions obtained with the NF and QRF models. Sources are assigned to tomographic bins using the posterior-median point estimate, \(z_{\rm median}\), predicted independently by each method. Consequently, the NF and QRF samples within the same nominal redshift interval may differ slightly (except the final tomographic bin which sample number are $\sim 12,000$ for NF and $\sim7,000$ for QRF ), and each reconstructed distribution is compared with the spectroscopic \(N(z)\) of its corresponding selected sample. For the NF, quoted uncertainties are the sample standard deviations across five independent training realisations using identical training and test samples but different random seeds and data shuffling. The QRF seed-to-seed variations are below the displayed numerical precision and are therefore not shown here.
}
\begin{tabular}{llccccc}
\toprule
Tomographic selection & Method &
$|\Delta z_{\rm mean}|$ ($\times10^{-3}$) &
$|\Delta\sigma|$ ($\times10^{-3}$) &
$W_1$ ($\times10^{-3}$) &
JS divergence ($\times10^{-3}$) &
Overlap($\%$) \\
\midrule

\multirow{2}{*}{$0<z_{\rm median}<0.2$}
& NF  & $2.6\pm1.1$ & $1.2\pm1.9$ & $3.0\pm1.1$ & $4.0\pm0.5$ & $93.1\pm0.4$ \\
& QRF & 4.2 & 7.9 & 5.1 & 7.4 & 90.7 \\

\addlinespace
\multirow{2}{*}{$0.2<z_{\rm median}<0.4$}
& NF  & $3.2\pm1.7$ & $0.6\pm0.4$ & $3.5\pm1.5$ & $1.3\pm0.2$ & $96.7\pm0.2$ \\
& QRF & 3.0 & 4.2 & 4.1 & 3.4 & 94.6 \\

\addlinespace
\multirow{2}{*}{$0.4<z_{\rm median}<0.6$}
& NF  & $3.6\pm1.8$ & $4.4\pm1.1$ & $4.7\pm1.3$ & $1.2\pm0.2$ & $96.6\pm0.3$ \\
& QRF & 3.0 & 12.7 & 9.7 & 5.6 & 92.6 \\

\addlinespace
\multirow{2}{*}{$0.6<z_{\rm median}<0.8$}
& NF  & $3.3\pm2.1$ & $6.6\pm1.2$ & $4.3\pm1.5$ & $1.1\pm0.2$ & $96.9\pm0.3$ \\
& QRF & 2.1 & 11.3 & 8.2 & 3.8 & 94.0 \\

\addlinespace
\multirow{2}{*}{$0.8<z_{\rm median}<1.0$}
& NF  & $1.6\pm0.5$ & $12.5\pm0.9$ & $6.5\pm0.8$ & $1.3\pm0.2$ & $97.0\pm0.5$ \\
& QRF & 4.3 & 14.7 & 9.8 & 2.6 & 94.7 \\

\addlinespace
\multirow{2}{*}{$1.0<z_{\rm median}<1.6$}
& NF  & $21\pm8$ & $5.3\pm7.3$ & $22\pm7$ & $1.2\pm0.3$ & $96.8\pm0.2$ \\
& QRF & 11.0 & 27.7 & 21.1 & 1.7 & 95.6 \\

\addlinespace
\multirow{2}{*}{$1.6<z_{\rm median}<2.5$}
& NF  & $18\pm14$ & $6.5\pm3.3$ & $21\pm11$ & $1.7\pm0.3$ & $96.1\pm0.5$ \\
& QRF & 20.7 & 26.1 & 26.8 & 2.3 & 95.1 \\

\addlinespace
\multirow{2}{*}{$z_{\rm median}>2.5$}
& NF  & $60\pm31$ & $30\pm20$ & $67\pm23$ & $11\pm2$ & $90.5\pm0.8$ \\
& QRF & 318 & 135 & 318 & 37.5 & 78.7 \\

\midrule
\multirow{2}{*}{Full sample}
& NF  & $2.8\pm3.3$ & $9.7\pm5.6$ & $7.8\pm1.9$ & $0.69\pm0.08$ & $97.5\pm0.2$ \\
& QRF & 0.6 & 3.6 & 4.5 & 0.73 & 97.2 \\

\bottomrule
\end{tabular}

\label{tab:Nzmetrics_bins}
\end{table*}

As shown in Table~\ref{tab:Nzmetrics_bins}, Fig.~\ref{fig:Nzthreebins}, and Fig.~\ref{fig:Nzhighbins}, the NF and QRF show different performance in recovering the mean redshift and the overall shape of the reconstructed \(N(z)\). Considering the central values, the NF yields smaller \(|\Delta\sigma|\) and JS divergence, and larger overlap, in all eight tomographic bins. The Wasserstein distance is also smaller for the NF in seven of the eight bins, with the only exception being \(1.0<z_{\rm median}<1.6\). In contrast, the mean-redshift bias shows a less systematic trend: QRF gives a smaller \(|\Delta z_{\rm mean}|\) in four bins, whereas the NF performs better in \(0<z_{\rm median}<0.2\), \(0.8<z_{\rm median}<1.0\), \(1.6<z_{\rm median}<2.5\) and \(z_{\rm median}>2.5\).

The largest difference between the NF and QRF is observed in the highest-redshift bin ($z_{\rm median}>2.5$). The NF achieves a smaller absolute mean-redshift bias ($|\Delta z_{\rm mean}|=0.060$) than the QRF ($0.318$), together with a lower Wasserstein distance ($W_1=0.067$ compared with $0.318$) and JS divergence ($0.011$ compared with $0.0375$). The overlap is also higher for the NF ($0.905$) than for the QRF ($0.787$). The high-redshift tomographic samples exhibit a sharp peak near $z_{\rm spec}=0$, dominated by stellar contaminants, as confirmed by their spectroscopic classifications. These contaminants are assigned high photometric redshift point estimates and consequently enter the high-$z$ tomographic samples, highlighting the limitations of point estimates in capturing redshift ambiguities. In contrast, the NF PDFs retain non-negligible probability at low redshifts, allowing PDF stacking to partially recover this component.

Differences are also evident at lower redshifts. For $0.4<z_{\rm median}<0.6$, the NF achieves a higher overlap ($0.966$ versus $0.926$) and a smaller Wasserstein distance ($4.7\times10^{-3}$ versus $9.7\times10^{-3}$) than the QRF. Overall, the NF provides a more accurate reconstruction of the tomographic $N(z)$, particularly at high redshifts, where photometric degeneracies are more significant and the available spectroscopic samples are smaller.

It is also evident from Figs.~\ref{fig:Nzthreebins} and \ref{fig:Nzhighbins} that the distributions constructed from the point estimates, \(N(z_{\rm median})\), do not reproduce the full shape of the spectroscopic \(N(z_{\rm spec})\), particularly at high redshift. A single point estimate cannot represent secondary redshift solutions, extended tails, or contamination associated with photometric degeneracies, all of which can remain encoded in the full conditional PDFs \(p(z\mid\mathbf{x})\).

For the full sample case, the differences between models are smaller. Without tomographic selection, QRF gives smaller \(|\Delta z_{\rm mean}|\), \(|\Delta\sigma|\), and \(W_1\), whereas the NF gives a slightly smaller JS divergence and a slightly larger overlap. Nevertheless, both methods recover the full-sample \(N(z)\) well, without obvious large-scale bias or leakage. This also shows that good recovery of the global \(N(z)\) does not necessarily imply equally accurate reconstruction after tomographic selection.

In summary, these results show that accurate reconstruction of tomographic \(N(z)\) benefits from retaining the full photo-\(z\) distribution, \(p(z\mid\mathbf{x})\), rather than relying only on point estimates. This is particularly important for cosmological tomography, where sources scattered across photo-\(z\) bin boundaries produce extended true-redshift tails that cannot be captured by \(z_{\rm median}\) alone. However, no calibration has yet been applied to the conditional photo-\(z\) PDFs considered here. Systematic miscalibration of individual PDFs may propagate through PDF stacking and affect the reconstructed \(N(z)\). We therefore assess the reliability of the predicted PDFs and investigate a practical calibration procedure in Section~\ref{sec:calibration}.

\subsection{PDF calibration and its impact on N(z) reconstruction} 
\label{sec:calibration}

In this section, we use the Probability Integral Transform (PIT)~\citep{Dawid1984PIT,Wittman2016PIT}, to assess the photo-$z$ PDFs predicted by the NF model. Motivated by the deviations from uniformity observed in the PIT distributions, we then apply a post-hoc cumulative distribution function (CDF) calibration following the approaches of~\citet{kuleshov2018accurate,Dey_etal2021}. Finally, we investigate whether the improved PDF estimation propagates to a more accurate reconstruction of the full and tomographic redshift distributions, $N(z)$.

For each galaxy with spectroscopic redshift $z_{{\rm spec},i}$, the PIT value is defined as:
\begin{equation}
    \mathrm{PIT}_i =
    \int_{0}^{z_{{\rm spec},i}} P(z\mid\mathbf{x}_i)\,{\rm d}z .
\end{equation}
If the PDFs are fully consistent with the distribution of the underlying spectroscopic sample, the PIT values should be uniformly distributed over \([0,1]\). A U-shaped distribution indicates over-confident PDFs, whereas an inverted-U shape indicates under-confidence; asymmetry and skewness generally reflect the existence of systematic bias. The PIT is particularly useful in this work because conventional photo-$z$ point-estimate metrics, such as the bias, $\sigma_{\rm NMAD}$,
and outlier fraction, compress each predicted distribution into a single value and cannot determine whether the full PDF is correctly estimated. Similarly, an accurate stacked $N(z)$ does not necessarily imply that the individual PDFs provide reliable probabilistic predictions. The PIT therefore provides a complementary test on the quality of the continuous NF PDFs estimation.

\begin{figure*}
    \centering
    \includegraphics[width=0.99\linewidth]{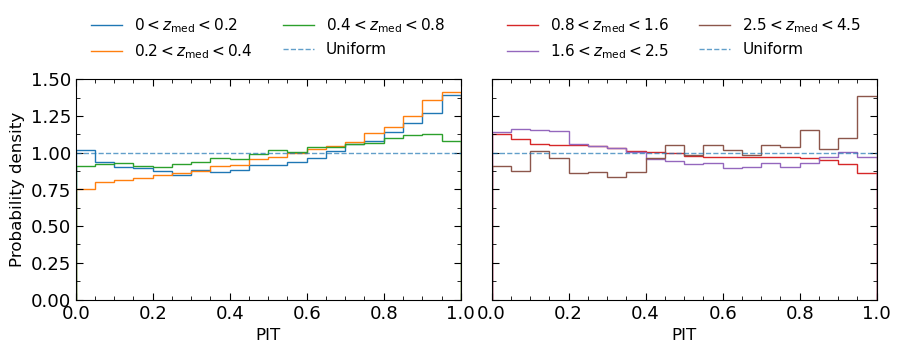}
    \caption{PIT distributions (20 bins) of the NF photo-$z$ PDFs in six tomographic bins defined by the NF-median redshifts, [0.0, 0.2, 0.4, 0.8, 1.6, 2.5, 4.5]. The dashed horizontal line indicates the uniform distribution. The strongest deviation from uniformity occurs at low redshift, with a deficit at low PIT values and an excess towards ${\rm PIT}=1$. The deviation becomes progressively weaker with increasing redshift, while the higher-redshift bins show more irregular variations across the PIT range. This redshift dependence motivates the use of separate calibration mappings in each tomographic redshift bin.}
    \label{fig:pit4bins}
\end{figure*}

Fig.~\ref{fig:pit4bins} shows the PIT distributions in six intervals of the posterior-median redshift, $z_{\rm median}$ with redshift bins [0.0, 0.2, 0.4, 0.8, 1.6, 2.5, 4.5]. The deviation from the uniform expectation varies clearly with redshift, with the asymmetric trend appearing in the redshift ranges of $0.2<z_{\rm{median}}<0.4$ and  $0.4<z_{\rm{median}}<0.8$, and the strongly irregular fluctuations in highest redshift range $z_{\rm{median}}>2.5$. The variations in PIT shape across redshift intervals motivate us to make the redshift-dependent post-hoc calibration introduced below. 

\begin{table*}
\centering
\caption{
Comparison of the tomographic and full $N(z)$ reconstruction before and after post-hoc calibration. Sources are divided into eight tomographic bins according to the posterior-median estimate, $z_{\rm median}$. The quoted values are the mean and sample standard deviations across five independent NF realisations. Absolute values are used for $\Delta z_{\rm mean}$ and $\Delta\sigma$.
}
\label{tab:calibration_tomographic}
\begin{tabular}{llccccc}
\hline
Tomographic selection &
Calibration &
$|\Delta z_{\rm mean}|(\times10^{-3})$ &
$|\Delta\sigma|(\times10^{-3})$ &
$W_1(\times10^{-3})$ &
JS divergence($\times10^{-3}$) &
Overlap($\%$)\\
\hline

\multirow{2}{*}{$0<z_{\rm median}<0.2$}
& Before
& $2.6 \pm 1.1$
& $1.2 \pm 1.9$
&$3.0 \pm 1.1$
& $4.0 \pm 0.5$
& $93.1 \pm 0.4$ \\

& After
& $2.5 \pm 1.8$
& $7.3\pm 2.6$
& $4.1 \pm 1.6$
& $4.4 \pm 0.8$
& $93.0 \pm 0.7$ \\
\addlinespace

\multirow{2}{*}{$0.2<z_{\rm median}<0.4$}
& Before
& $3.2 \pm 1.7$
& $0.6 \pm 0.4$
& $3.5 \pm 1.5$
& $1.3\pm 0.2$
& $96.7 \pm 0.2$ \\

& After
& $1.6 \pm 1.0$
& $1.0 \pm 0.5$
& $2.3 \pm 0.8$
& $1.1 \pm 0.2$
& $97.0 \pm 0.2$ \\
\addlinespace

\multirow{2}{*}{$0.4<z_{\rm median}<0.6$}
& Before
& $3.6 \pm 1.8$
& $4.4 \pm 1.1$
& $4.7 \pm 1.3$
& $1.2 \pm 0.2$
& $96.6 \pm 0.3$ \\

& After
& $1.8 \pm 1.5$
& $4.0 \pm 1.6$
& $3.5\pm 1.3$
& $1.1 \pm 0.2$
& $96.8 \pm 0.3$ \\
\addlinespace

\multirow{2}{*}{$0.6<z_{\rm median}<0.8$}
& Before
& $3.3 \pm 2.1$
& $ 6.6\pm 1.2$
& $4.3 \pm 1.5$
& $1.1 \pm 0.2$
& $96.9 \pm 0.3$ \\

& After
& $1.8 \pm 0.9$
& $4.6 \pm 0.9$
& $3.1 \pm 0.3$
& $0.90 \pm 0.08$
& $97.2 \pm 0.1$ \\
\addlinespace
\multirow{2}{*}{$0.8<z_{\rm median}<1.0$}
& Before
& $1.6 \pm 0.5$
& $12.5\pm0. 9$
& $6.5 \pm 0.8$
& $1.3 \pm 0.2$
& $97.0 \pm 0.5$ \\

& After
& $1.9 \pm 1.8$
& $8.6 \pm 1.8$
& $6.5 \pm 0.8$
& $1.2 \pm 0.2$
& $96.8 \pm 0.5$ \\
\addlinespace

\multirow{2}{*}{$1.0<z_{\rm median}<1.6$}
& Before
& $21 \pm 8$
& $5.3\pm 7.3$
& $22 \pm 7$
& $1.2 \pm 0.3$
& $96.8 \pm 0.2$ \\

& After
& $26 \pm 8$
& $8.4 \pm 4.8$
& $28 \pm 6$
& $1.7 \pm 0.3$
& $96.4 \pm 0.2$ \\
\addlinespace

\multirow{2}{*}{$1.6<z_{\rm median}<2.5$}
& Before
& $18 \pm 14$
& $6.5\pm 3.3$
& $21 \pm 11$
& $1.7\pm 0.3$
& $96.1 \pm 0.5$ \\

& After
& $11 \pm 6$
& $18 \pm 6$
& $17 \pm 2$
& $1.6 \pm 0.2$
& $96.2 \pm 0.2$ \\
\addlinespace
\multirow{2}{*}{$z_{\rm median}>2.5$}
& Before
& $60 \pm 31$
& $30\pm 20$
& $67 \pm 23$
& $11\pm 2$
& $90.5\pm 0.8$ \\

& After
& $16 \pm 11$
& $28 \pm 26$
& $35 \pm 7$
& $11 \pm 2$
& $90.7 \pm1.0$ \\
\addlinespace
\hline

Full Samples
& Before
& $2.8 \pm 3.3$
& $9.7 \pm 5.6$
& $7.8 \pm 1.9$
& $0.69 \pm 0.08$
& $97.5 \pm 0.2$ \\

& After
& $4.6 \pm 3.6$
& $8.1 \pm 3.6$
& $7.6 \pm 1.1$
& $0.64 \pm 0.07$
& $97.5 \pm 0.1$ \\
\hline
\end{tabular}
\end{table*}

To correct the PIT deviations shown in
Fig.~\ref{fig:pit4bins}, we first tested global
temperature scaling~\citep{guo2017calibration},
\begin{equation}
p_T(z\mid\mathbf{x})
\propto
p(z\mid\mathbf{x})^{1/T}.
\end{equation}
However, a single temperature parameter $T$ could not
adequately correct the observed PIT deviations,
indicating that global broadening or narrowing
of the PDFs is insufficient. We therefore adopt
a more flexible empirical CDF calibration method,
following \citet{kuleshov2018accurate,Dey_etal2021},
using the independent validation sample described
in Section~\ref{subsec:data}.  The core idea of this method is to construct a mapping bringing the estimated PIT distribution to the uniform distribution. This mapping is then applied to rescale the estimated PDFs.

The calibration is performed separately in six redshift bins defined by the predicted median redshift, $z_{\rm median}$, with boundaries $[0.0, 0.2, 0.4, 0.8, 1.6, 2.5, 4.5]$. For each galaxy in the validation sample, we first compute the PIT value
\begin{equation}
u_i = F_i(z_{{\rm spec},i}),
\end{equation}

where $F_i(z)$ is the CDF predicted by the NF model. For each redshift bin $b$, we estimate the empirical CDF of the PIT values, denoted by $G_b(u)$.

To construct the calibration mapping, we define 101 equally spaced quantile levels,
\begin{equation}
q_k = \frac{k}{100}, \qquad k=0,\ldots,100,
\end{equation}

and calculate the corresponding empirical PIT quantiles, $u_{b,k}$, from the validation sample in each redshift bin. The boundary values are explicitly set to $u_{b,0}=0$ and $u_{b,100}=1$. The calibration function $G_b(u)$ is then approximated by linear interpolation between adjacent knots $(u_{b,k},q_k)$, providing a continuous mapping from the original PIT values to calibrated quantile levels. The calibrated predictive CDF for galaxy $i$ is defined as

\begin{equation}
F_{i,\rm cal}(z)=G_b\left[F_i(z)\right],
\end{equation}

where $b$ denotes the redshift bin assigned to the galaxy based on its original predicted median redshift.

In practice, the original predictive distribution of each galaxy is represented by 200 Monte Carlo redshift samples generated by the NF model. The calibration is implemented through quantile remapping. For each uniformly spaced quantile level $q_j$ of the calibrated distribution, the corresponding quantile level of the original distribution is obtained through the inverse calibration mapping, $G_b^{-1}(q_j)$. The calibrated redshift values are then calculated as
\begin{equation}
z_{i,j}^{\rm cal}
=
\widehat F_i^{-1}\left[G_b^{-1}(q_j)\right],
\qquad
q_j=\frac{j-1/2}{200},
\quad j=1,\ldots,200,    
\end{equation}

where $\widehat F_i^{-1}$ denotes the empirical quantile function estimated from the original Monte Carlo samples. This procedure modifies the shapes of the predictive redshift distributions while preserving the number of samples used for subsequent analyses. The calibration mappings are derived exclusively from the validation sample and subsequently applied to the independent testing sample.

We note that deriving the calibration mappings directly from the testing sample can generally lead to greater improvements in the tomographic $N(z)$ reconstruction metrics. However, such improvements are not necessarily consistent across all tomographic bins and metrics. Calibrating and evaluating on the same testing sample introduces data leakage and may lead to overly optimistic estimates of calibration performance. Such an approach is also not representative of realistic cosmological analysis pipelines, where spectroscopic redshifts are generally unavailable for the target photometric sample. We therefore adopt validation-based calibration to provide a more statistically reliable assessment of the method's generalisation to independent data.

The calibration mappings are constructed exclusively from the spectroscopic validation sample of 121,594 sources (Section~\ref{sec:data}) and applied to the spatially independent testing sample. This ensures an out-of-sample evaluation, with testing $z_{\rm spec}$ values used only to assess the calibrated PIT distributions and $N(z)$ reconstruction. For each NF realisation, a separate mapping is derived from its validation-set PIT distribution and applied to the corresponding testing PDFs.

\begin{figure}
    \centering
    \includegraphics[width=0.90\linewidth]{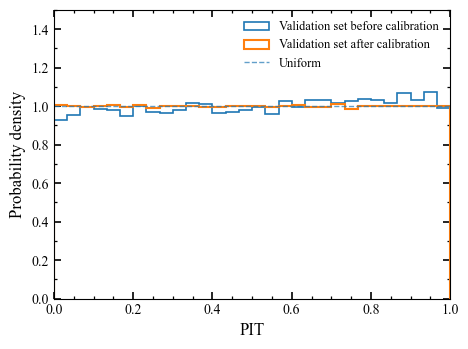}
     \includegraphics[width=0.90\linewidth]{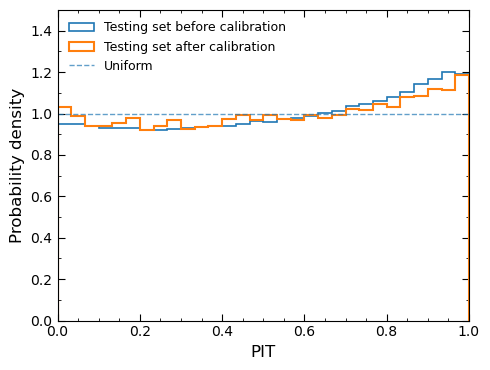}
    \caption{PIT distributions (30 bins) for the independent validation sample (upper panel) and testing sample (lower panel), shown before and after applying the calibration mapping derived from the validation set. The dashed horizontal line marks the ideal uniform distribution. After calibration, the validation-sample PIT is close to uniform, while the testing-sample PIT shows only modest improvement, with residual excesses remaining.}
    \label{fig:pitcalibration}
\end{figure}

\begin{figure}
    \centering
    \includegraphics[width=0.95\linewidth]{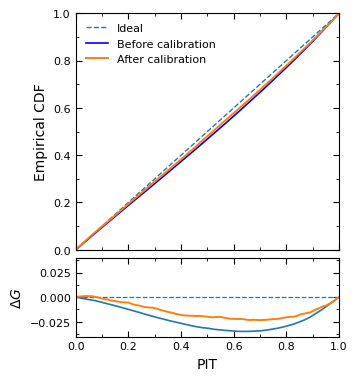}
    \caption{Empirical Cumulative Distribution Functions (CDF) of the PIT values for the independent testing sample before and after calibration. The dashed diagonal shows the ideal relation expected for uniformly distributed PIT values. The lower panel shows the residual, $\Delta G(u)=G(u)-u$, relative to the ideal CDF. After calibration, the empirical CDF moves closer to the uniform expectation and the magnitude of the residual is reduced over most of the PIT range, indicating improved out-of-sample probabilistic calibration.
}
    \label{fig:ECDFcalibration}
\end{figure}

Fig.~\ref{fig:pitcalibration} shows the PIT distributions before and after calibration. As expected, the validation PIT becomes approximately uniform by construction, while the testing PIT also shows reduced deviations from uniformity, indicating that the calibration transfers to the independent sample. We further quantify the calibration using the empirical CDF, $G(u)$, and its residual, $\Delta G(u)=G(u)-u$, shown in Fig.~\ref{fig:ECDFcalibration}. For perfectly calibrated PDFs, $G(u)=u$,  deviations from the identity relation directly indicate cumulative miscalibration.

We then use the calibrated NF PDFs to reconstruct the redshift distributions $N(z)$. Table~\ref{tab:calibration_tomographic} compares the reconstruction performance before and after calibration. At intermediate redshifts ($0.2<z_{\rm median}<0.8$), the absolute mean-redshift bias decreases to $(1.6$--$1.8)\times10^{-3}$, alongside reductions in the Wasserstein distance and JS divergence. The largest improvement occurs at $z_{\rm median}>2.5$, where the bias decreases from $0.060$ to $0.016$ and the Wasserstein distance from $0.067$ to $0.035$. However, calibration increases the distribution-width discrepancies in some bins and slightly worsens the full-sample mean-redshift bias, while the overlap remains broadly unchanged.

\begin{figure}
    \centering
    \includegraphics[width=0.95\linewidth]{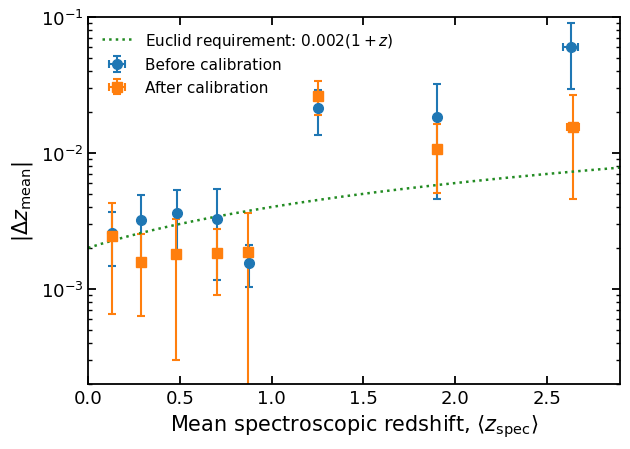}
     \caption{
Absolute mean-redshift bias,
$|\Delta z_{\rm mean}| =
|\langle z_{\rm pred}\rangle-\langle z_{\rm spec}\rangle|$,
as a function of the mean spectroscopic redshift for eight tomographic bins before (blue circles) and after (orange squares) post-hoc calibration.
The points represent the mean absolute biases over independent NF realisations, with error bars indicating the sample standard deviations.
The green dotted line indicates the Euclid mean-redshift accuracy benchmark, $0.002(1+z)$.
Tomographic bins are defined using the posterior-median redshift, $z_{\rm median}$, and galaxy membership is recomputed after calibration, accounting for the small differences in the mean spectroscopic redshifts before and after calibration.
The lower error bar for the fifth tomographic bin extends below the plotted lower limit and is therefore partially truncated on the logarithmic $y$-axis.
}
\label{fig:deltazmean}
\end{figure}

Fig.~\ref{fig:deltazmean} compares the absolute
mean-redshift bias with the Euclid weak-lensing
requirement, $|\Delta z_{\rm mean}|<0.002(1+z)$
~\citep{EuclidNz,Euclid2021Photoz}, where $z$
is taken as the mean spectroscopic redshift
of each tomographic bin.
After calibration, the bias in the first bin
decreases slightly from 0.0026 to 0.0025,
approaching but remaining above the Euclid
benchmark. The second to fifth bins
($0.2<z_{\rm median}<1.0$) satisfy the benchmark
after calibration.
At higher redshifts, the bias increases from
0.021 to 0.026 for $1.0<z_{\rm median}<1.6$,
but decreases from 0.018 to 0.011 for
$1.6<z_{\rm median}<2.5$ and from 0.060
to 0.016 for $z_{\rm median}>2.5$.
Nevertheless, all three higher-redshift bins
remain above the Euclid benchmark.

We emphasise that satisfying this benchmark
within the DESI-selected spectroscopic sample
does not demonstrate that the same accuracy
can be achieved for Euclid weak-lensing source
galaxies, which have different selection
functions and photometric properties.
Further validation on representative target
populations is therefore required to assess
the applicability of the framework to Euclid.

Generally speaking, redshift-dependent post-hoc calibration improves the probabilistic consistency of NF photo-z PDFs and leads to modest improvements in the reconstructed $N(z)$. The validation-derived mapping brings both the PIT distribution and its empirical CDF closer to the uniform expectation when applied to the independent test sample, although residual miscalibration remains. For the full sample, the Wasserstein distance and JS divergence decrease slightly, while the mean-redshift bias increases. The tomographic results show more substantial improvements, particularly at low and intermediate redshifts, where the mean-redshift biases in the second to fifth bins ($0.2<z_{\rm{median}}<1.0$) fall within the Euclid accuracy benchmark after calibration. At higher redshifts, the improvements are less consistent, although substantial reductions in the mean-redshift bias are observed in selected bins. The incomplete transfer of calibration may partly reflect differences between the validation and test samples, which occupy different sky regions. Nevertheless, the improvements achieved without using spectroscopic redshifts from the test sample highlight the potential of PIT-based recalibration for improving probabilistic photo-$z$ estimates and tomographic redshift distributions in different catalogues.

\section{Discussion}
\label{sec:discuss}
\subsection{PDF calibration}

Accurate calibration of predicted redshift PDFs is essential for reliable reconstruction of the ensemble redshift distribution, $N(z)$. As shown in Section~\ref{sec:calibration}, empirical CDF calibration~\citep{kuleshov2018accurate,Dey_etal2021} produces an approximately uniform PIT distribution for the validation sample, while modest deviations remain when the mapping is applied to the independent testing sample. Calibration also improves $|\Delta z_{\rm mean}|$ in several tomographic bins. However, the validation and testing samples exhibit similar photometric features and redshift distributions (Fig.~\ref{fig:featuredistributions}), leaving the effectiveness of this approach for more distinct galaxy populations uncertain.

Compared with simpler approaches such as temperature scaling~\citep{guo2017calibration}, empirical CDF recalibration with 101 interpolation knots provides greater flexibility in correcting complex PIT deviations. However, the resulting mapping is determined by the PIT distribution of the calibration sample and is therefore not necessarily universal. Even an accurately determined calibration mapping may not remain valid for a target population with different photometric properties or redshift distributions. This limitation is particularly relevant to practical photo-$z$ applications, where representative spectroscopic samples are often unavailable for the target population. 

Therefore, the key challenge is not simply to achieve a uniform PIT distribution on the validation sample, but to develop calibration methods that remain reliable across different galaxy populations. Future work could explore alternative post-hoc calibration approaches, such as isotonic regression~\citep{niculescu2005predicting}, and methods that account for photometric features or individual PDF properties, such as non-Gaussianity and multimodality, while evaluating their transferability using independent datasets. Importantly, achieving PIT uniformity does not necessarily guarantee accurate reconstruction of tomographic $N(z)$, further motivating calibration methods that are both transferable across galaxy populations and reliable for distribution-level cosmological analyses~\citep{rau2015accurate,EuclidNz,Euclid2021Photoz}.

\subsection{Characteristics of NF-based photo-\(z\) estimation}

In our tomographic analysis, we find that the NF results generally outperform the QRF ones, whereas the advantage is less pronounced for the full sample. These findings, together with the point estimate results, are consistent with the interpretation that NF photo-$z$ estimation excels when the data are noisy and sparse \citep{ren2026NF}.

Our analysis also highlights several practical characteristics of NF-based photo-\(z\) inference. First, repeated training runs show noticeable run-to-run variations even when the same training and testing samples are used (see Section~\ref{sec:tomoN(z)}). This training-induced variability introduces an additional source of uncertainty beyond the systematic effects associated with the data. Further investigation of convergence, optimization, model architecture, and initialization is therefore required to determine whether this variability can be substantially reduced. 

At the same time, the continuous probabilistic representation provided by NFs offers useful flexibility for PDF-based analyses. For diagnostics such as PIT, the NF provides an explicit continuous predictive distribution, while individual PDFs can be sampled at arbitrarily fine numerical resolution without retraining the model. For the smaller high-redshift sample, the reduced computational cost allows us to draw 1,000 or more Monte Carlo samples per source. Hence, the continuous formulation of the NF is convenient for probabilistic diagnostics and individual-PDF studies.

\subsection{Limitations and future work}

Although the training and testing samples occupy
non-overlapping sky regions, both are constructed
from the DESI spectroscopic sample using the same
photometric and quality selections from DESI Legacy DR10. As the primary
goal of this study is to validate the probabilistic
photo-$z$ estimation and $N(z)$ reconstruction
methodology, this controlled setting allows us to
assess the performance of the NF model without
introducing additional differences in sample selection.

However, this experimental design does not fully
investigate its generalisability to broader photometric
populations with different distributions of colours,
magnitudes, or galaxy types. In particular, the
tomographic $N(z)$ accuracy achieved in this work
should not be interpreted as the expected performance
for Euclid weak-lensing source galaxies.
Recent studies have highlighted the importance of
spectroscopic incompleteness and sample
representativeness for redshift calibration
\citep{Lange2026,Blanco2026}.

Nevertheless, the NF framework itself is not specific
to DESI but offers considerable flexibility in
modelling conditional redshift distributions.
It can be adapted to different photometric surveys,
including Euclid, LSST and CSST by incorporating their
respective photometric features and retraining or
recalibrating the model using suitable reference data.
The method may also be particularly promising for
lens galaxy samples, whose photometric properties
can be more similar to those of DESI spectroscopic
targets. Future applications will require careful
assessment of selection effects and differences
between the spectroscopic training and target
photometric populations.

\section{Conclusions}
\label{sec:conclusion}

In this work, we investigate conditional Normalizing Flows (NFs) for probabilistic photometric-redshift estimation and redshift-distribution reconstruction using DESI DR1 spectroscopic data and DESI Legacy Imaging Surveys photometry. The main analysis uses an independent test sample of 2,028,371 sources spanning around $0<z_{\rm spec}<4.5$, with the NF evaluated over five independent training realisations and compared with the Random Forest (RF) and Quantile Random Forest (QRF) models. For point estimation, the NF achieves lower $\sigma_{\rm NMAD}$ and outlier fractions compare to RF and QRF across the full redshift range, although its bias is slightly larger than that of QRF and exhibits greater run-to-run variation. At higher redshifts ($z_{\rm spec}>1.6$), where the performance of all three models generally deteriorates due to the limited samples' number at high redshift (see Fig.~\ref{fig:Nztraining}), the NF demonstrates a clearer advantage in bias, scatter, and outlier fraction, consistent with \citet{ren2026NF}. However, point-estimation metrics alone do not fully capture the advantages of the NF, which provides continuous conditional redshift distributions that retain information about their widths, asymmetries, and extended tails.

This advantage becomes more apparent in tomographic $N(z)$ reconstruction. Across the first seven tomographic bins, the NF generally recovers the spectroscopic redshift distributions more accurately than QRF in terms of distribution width, Wasserstein distance, JS divergence, and probability overlap, although QRF achieves smaller mean-redshift biases in some bins. In the highest-redshift bin ($z_{\rm median}>2.5$), the NF shows a particularly substantial improvement over QRF across the reconstruction metrics, including its ability to recover the extended low-redshift tail. For the full-sample $N(z)$, both probabilistic methods perform well, indicating that the advantages of the NF are more pronounced in tomographic reconstruction than in global distribution-level statistics.

We further evaluate the probabilistic calibration of the NF PDFs using the probability integral transform (PIT) and apply a redshift-dependent empirical CDF recalibration derived from an independent validation sample. When transferred to the spatially independent test sample, the calibrated PIT distributions move closer to the uniform expectation, although residual miscalibration remains. The recalibration also leads to improvements in tomographic $N(z)$ reconstruction in several redshift bins, particularly in the mean-redshift accuracy, while largely preserving the overall distribution shapes. After calibration, the mean absolute redshift biases in the second through fifth tomographic bins at $0.2<z_{\rm median}<1$ fall below the Euclid accuracy benchmark of $0.002(1+z)$. At higher redshifts, the results are less consistent, although substantial reductions in the mean-redshift bias are achieved for $1.6<z_{\rm median}<2.5$ and $z_{\rm median}>2.5$. These findings demonstrate the flexibility of continuous NF PDFs for post-hoc calibration, while also showing that improved probabilistic consistency does not necessarily translate into better $N(z)$ reconstruction across all metrics.

Collectively, our results demonstrate that the
advantages of Normalizing Flows for photometric-redshift
inference extend beyond point-estimation accuracy
to their flexible representation of the full
conditional redshift probability distribution.
Retaining this probabilistic information can improve
tomographic $N(z)$ reconstruction, particularly
when broad or multimodal redshift distributions
lead to substantial leakage across photo-$z$
bin boundaries. Nevertheless, the observed
run-to-run variability, the uncertain generalisation
to different photometric populations, and the
limitations of post-hoc PDF calibration remain
important challenges. Future work should explore
more transferable calibration strategies, evaluate
NFs on different target populations and against
a broader range of probabilistic photo-$z$ methods,
and incorporate the reconstructed redshift
distributions into galaxy-clustering and
weak-lensing likelihood analyses to quantify
their impact on cosmological inference.

\section*{Acknowledgements}

The authors would like to thank Prof. Marc Manera (IFAE) for his valuable suggestions on the reconstruction of redshift distributions, $N(z)$, and for his helpful comments on the Introduction and Willian d'Assignies for his valuable comments on the manuscript. ZC and KCC acknowledge the support by the National Natural Science Foundation of China under the grant number 12273121 and 12533002, and the science research grants from the China Manned Space Project with NO.CMS-CSST-2021-B01.  AR acknowledges support from the Swiss National Science Foundation (SNF) “Cosmology with 3D Maps of the Universe” research grant 200020\_207379. 

This research used data obtained with the Dark Energy Spectroscopic Instrument (DESI). DESI construction and operations is managed by the Lawrence Berkeley National Laboratory. This material is based upon work supported by the U.S. Department of Energy, Office of Science, Office of High-Energy Physics, under Contract No. DE–AC02–05CH11231, and by the National Energy Research Scientific Computing Center, a DOE Office of Science User Facility under the same contract. Additional support for DESI was provided by the U.S. National Science Foundation (NSF), Division of Astronomical Sciences under Contract No. AST-0950945 to the NSF’s National Optical-Infrared Astronomy Research Laboratory; the Science and Technology Facilities Council of the United Kingdom; the Gordon and Betty Moore Foundation; the Heising-Simons Foundation; the French Alternative Energies and Atomic Energy Commission (CEA); the National Council of Humanities, Science and Technology of Mexico (CONAHCYT); the Ministry of Science and Innovation of Spain (MICINN), and by the DESI Member Institutions: www.desi.lbl.gov/collaborating-institutions. The DESI collaboration is honored to be permitted to conduct scientific research on I’oligam Du’ag (Kitt Peak), a mountain with particular significance to the Tohono O’odham Nation. Any opinions, findings, and conclusions or recommendations expressed in this material are those of the author(s) and do not necessarily reflect the views of the U.S. National Science Foundation, the U.S. Department of Energy, or any of the listed funding agencies. The DESI Legacy Imaging Surveys consist of three individual and complementary projects: the Dark Energy Camera Legacy Survey (DECaLS), the Beijing-Arizona Sky Survey (BASS), and the Mayall z-band Legacy Survey (MzLS). DECaLS, BASS and MzLS together include data obtained, respectively, at the Blanco telescope, Cerro Tololo Inter-American Observatory, NSF’s NOIRLab; the Bok telescope, Steward Observatory, University of Arizona; and the Mayall telescope, Kitt Peak National Observatory, NOIRLab. NOIRLab is operated by the Association of Universities for Research in Astronomy (AURA) under a cooperative agreement with the National Science Foundation. Pipeline processing and analyses of the data were supported by NOIRLab and the Lawrence Berkeley National Laboratory (LBNL). Legacy Surveys also uses data products from the Near-Earth Object Wide-field Infrared Survey Explorer (NEOWISE), a project of the Jet Propulsion Laboratory/California Institute of Technology, funded by the National Aeronautics and Space Administration. Legacy Surveys was supported by: the Director, Office of Science, Office of High Energy Physics of the U.S. Department of Energy; the National Energy Research Scientific Computing Center, a DOE Office of Science User Facility; the U.S. National Science Foundation, Division of Astronomical Sciences; the National Astronomical Observatories of China, the Chinese Academy of Sciences and the Chinese National Natural Science Foundation. LBNL is managed by the Regents of the University of California under contract to the U.S. Department of Energy. The complete acknowledgments can be found at https://www.legacysurvey.org/acknowledgment/.


\section*{Data Availability}

The spectroscopic data used in this work are publicly available from the DESI Data Release 1 (DR1) at https://data.desi.lbl.gov/public/dr1/. The photometric data are publicly available from the DESI Legacy Imaging Surveys Data Release 10 (DR10) at https://www.legacysurvey.org/dr10/.

The training, validation, and testing catalogues constructed by cross-matching these datasets and applying the selection criteria described in Section 2 are available on Zenodo at https://doi.org/10.5281/zenodo.23642619.

The source code used for model training, photometric-redshift estimation, and redshift-distribution reconstruction, together with the five trained NF models, is publicly available at https://github.com/chenzhuo0623-pixel/Normalizing-flow-photometric-redshift-PDFs-with-DESI-DR1.




\bibliographystyle{mnras}
\bibliography{reference} 

@ARTICLE{Hildebrandt_etal2021,
       author = {{Hildebrandt}, H. and {van den Busch}, J.~L. and {Wright}, A.~H. and {Blake}, C. and {Joachimi}, B. and {Kuijken}, K. and {Tr{\"o}ster}, T. and {Asgari}, M. and {Bilicki}, M. and {de Jong}, J.~T.~A. and {Dvornik}, A. and {Erben}, T. and {Getman}, F. and {Giblin}, B. and {Heymans}, C. and {Kannawadi}, A. and {Lin}, C. -A. and {Shan}, H. -Y.},
        title = "{KiDS-1000 catalogue: Redshift distributions and their calibration}",
      journal = {\aap},
         year = 2021,
        month = mar,
       volume = {647},
          eid = {A124},
        pages = {A124},
          doi = {10.1051/0004-6361/202039018},
archivePrefix = {arXiv},
       eprint = {2007.15635},
 primaryClass = {astro-ph.CO},
       adsurl = {https://ui.adsabs.harvard.edu/abs/2021A&A...647A.124H}
}

@ARTICLE{Rau_etal2023,
       author = {{Rau}, Markus Michael and {Dalal}, Roohi and {Zhang}, Tianqing and {Li}, Xiangchong and {Nishizawa}, Atsushi J. and {More}, Surhud and {Mandelbaum}, Rachel and {Miyatake}, Hironao and {Strauss}, Michael A. and {Takada}, Masahiro},
        title = "{Weak lensing tomographic redshift distribution inference for the Hyper Suprime-Cam Subaru Strategic Program three-year shape catalogue}",
      journal = {\mnras},
         year = 2023,
        month = oct,
       volume = {524},
       number = {4},
        pages = {5109-5131},
          doi = {10.1093/mnras/stad1962},
archivePrefix = {arXiv},
       eprint = {2211.16516},
 primaryClass = {astro-ph.CO},
       adsurl = {https://ui.adsabs.harvard.edu/abs/2023MNRAS.524.5109R}
}

@article{Cawthon_etal2022,
	adsurl = {https://ui.adsabs.harvard.edu/abs/2022MNRAS.513.5517C},
	archiveprefix = {arXiv},
	author = {{Cawthon}, R. and others},
	doi = {10.1093/mnras/stac1160},
	eprint = {2012.12826},
	journal = {\mnras},
	month = jul,
	number = {4},
	pages = {5517-5539},
	primaryclass = {astro-ph.CO},
	title = {{Dark Energy Survey Year 3 results: calibration of lens sample redshift distributions using clustering redshifts with BOSS/eBOSS}},
	volume = {513},
	year = 2022}

@ARTICLE{Gatti_etal2018,
       author = {{Gatti}, M. and {Vielzeuf}, P. and {Davis}, C. and {Cawthon}, R. and {Rau}, M.~M. and {DeRose}, J. and {De Vicente}, J. and {Alarcon}, A. and {Rozo}, E. and {Gaztanaga}, E. and {Hoyle}, B. and {Miquel}, R. and {Bernstein}, G.~M. and {Bonnett}, C. and {Carnero Rosell}, A. and {Castander}, F.~J. and {Chang}, C. and {da Costa}, L.~N. and {Gruen}, D. and {Gschwend}, J. and {Hartley}, W.~G. and {Lin}, H. and {MacCrann}, N. and {Maia}, M.~A.~G. and {Ogando}, R.~L.~C. and {Roodman}, A. and {Sevilla-Noarbe}, I. and {Troxel}, M.~A. and {Wechsler}, R.~H. and {Asorey}, J. and {Davis}, T.~M. and {Glazebrook}, K. and {Hinton}, S.~R. and {Lewis}, G. and {Lidman}, C. and {Macaulay}, E. and {M{\"o}ller}, A. and {O'Neill}, C.~R. and {Sommer}, N.~E. and {Uddin}, S.~A. and {Yuan}, F. and {Zhang}, B. and {Abbott}, T.~M.~C. and {Allam}, S. and {Annis}, J. and {Bechtol}, K. and {Brooks}, D. and {Burke}, D.~L. and {Carollo}, D. and {Carrasco Kind}, M. and {Carretero}, J. and {Cunha}, C.~E. and {D'Andrea}, C.~B. and {DePoy}, D.~L. and {Desai}, S. and {Eifler}, T.~F. and {Evrard}, A.~E. and {Flaugher}, B. and {Fosalba}, P. and {Frieman}, J. and {Garc{\'\i}a-Bellido}, J. and {Gerdes}, D.~W. and {Goldstein}, D.~A. and {Gruendl}, R.~A. and {Gutierrez}, G. and {Honscheid}, K. and {Hoormann}, J.~K. and {Jain}, B. and {James}, D.~J. and {Jarvis}, M. and {Jeltema}, T. and {Johnson}, M.~W.~G. and {Johnson}, M.~D. and {Krause}, E. and {Kuehn}, K. and {Kuhlmann}, S. and {Kuropatkin}, N. and {Li}, T.~S. and {Lima}, M. and {Marshall}, J.~L. and {Melchior}, P. and {Menanteau}, F. and {Nichol}, R.~C. and {Nord}, B. and {Plazas}, A.~A. and {Reil}, K. and {Rykoff}, E.~S. and {Sako}, M. and {Sanchez}, E. and {Scarpine}, V. and {Schubnell}, M. and {Sheldon}, E. and {Smith}, M. and {Smith}, R.~C. and {Soares-Santos}, M. and {Sobreira}, F. and {Suchyta}, E. and {Swanson}, M.~E.~C. and {Tarle}, G. and {Thomas}, D. and {Tucker}, B.~E. and {Tucker}, D.~L. and {Vikram}, V. and {Walker}, A.~R. and {Weller}, J. and {Wester}, W. and {Wolf}, R.~C.},
        title = "{Dark Energy Survey Year 1 results: cross-correlation redshifts - methods and systematics characterization}",
      journal = {\mnras},
         year = 2018,
        month = jun,
       volume = {477},
       number = {2},
        pages = {1664-1682},
          doi = {10.1093/mnras/sty466},
archivePrefix = {arXiv},
       eprint = {1709.00992},
 primaryClass = {astro-ph.CO},
       adsurl = {https://ui.adsabs.harvard.edu/abs/2018MNRAS.477.1664G}
}

@article{DESI2022,
  author = {{DESI Collaboration}},
  title = {The Dark Energy Spectroscopic Instrument (DESI)},
  journal = {Astronomical Journal},
  year = {2022},
  volume = {164},
  pages = {207}
}

@article{Desi2019,
  author = {Dey, A. and others},
  title = {Overview of the DESI Legacy Imaging Surveys},
  journal = {Astronomical Journal},
  year = {2019},
  volume = {157},
  pages = {168}
}

@article{DESI2026DR1,
    author  = {{DESI Collaboration}},
    title   = {Data Release 1 of the Dark Energy Spectroscopic Instrument},
    journal = {AJ},
    year    = {2026},
    volume  = {171},
    number  = {5},
    pages   = {285},
    doi     = {10.3847/1538-3881/ae4c43}
}

@article{Wright2010WISE,
  author  = {Wright, Edward L. and Eisenhardt, Peter R. M. and Mainzer, Amy K. and others},
  title   = {The Wide-field Infrared Survey Explorer (WISE): Mission Description and Initial On-orbit Performance},
  journal = {The Astronomical Journal},
  year    = {2010},
  volume  = {140},
  number  = {6},
  pages   = {1868--1881},
  doi     = {10.1088/0004-6256/140/6/1868},
  url     = {https://iopscience.iop.org/article/10.1088/0004-6256/140/6/1868}
}

@article{Flaugher2015DECam,
  author  = {Flaugher, Brenna and Diehl, H. Thomas and Honscheid, Klaus and others},
  title   = {The Dark Energy Camera},
  journal = {The Astronomical Journal},
  year    = {2015},
  volume  = {150},
  number  = {5},
  pages   = {150},
  doi     = {10.1088/0004-6256/150/5/150},
  url     = {https://iopscience.iop.org/article/10.1088/0004-6256/150/5/150}
}

@ARTICLE{Euclid_2011,
       author = {{Laureijs}, R. and {Amiaux}, J. and {Arduini}, S. and {Augu{\`e}res}, J. -L. and {Brinchmann}, J. and {Cole}, R. and {Cropper}, M. and {Dabin}, C. and {Duvet}, L. and {Ealet}, A. and {Garilli}, B. and {Gondoin}, P. and {Guzzo}, L. and {Hoar}, J. and {Hoekstra}, H. and {Holmes}, R. and {Kitching}, T. and {Maciaszek}, T. and {Mellier}, Y. and {Pasian}, F. and {Percival}, W. and {Rhodes}, J. and {Saavedra Criado}, G. and {Sauvage}, M. and {Scaramella}, R. and {Valenziano}, L. and {Warren}, S. and {Bender}, R. and {Castander}, F. and {Cimatti}, A. and {Le F{\`e}vre}, O. and {Kurki-Suonio}, H. and {Levi}, M. and {Lilje}, P. and {Meylan}, G. and {Nichol}, R. and {Pedersen}, K. and {Popa}, V. and {Rebolo Lopez}, R. and {Rix}, H. -W. and {Rottgering}, H. and {Zeilinger}, W. and {Grupp}, F. and {Hudelot}, P. and {Massey}, R. and {Meneghetti}, M. and {Miller}, L. and {Paltani}, S. and {Paulin-Henriksson}, S. and {Pires}, S. and {Saxton}, C. and {Schrabback}, T. and {Seidel}, G. and {Walsh}, J. and {Aghanim}, N. and {Amendola}, L. and {Bartlett}, J. and {Baccigalupi}, C. and {Beaulieu}, J. -P. and {Benabed}, K. and {Cuby}, J. -G. and {Elbaz}, D. and {Fosalba}, P. and {Gavazzi}, G. and {Helmi}, A. and {Hook}, I. and {Irwin}, M. and {Kneib}, J. -P. and {Kunz}, M. and {Mannucci}, F. and {Moscardini}, L. and {Tao}, C. and {Teyssier}, R. and {Weller}, J. and {Zamorani}, G. and {Zapatero Osorio}, M.~R. and {Boulade}, O. and {Foumond}, J.~J. and {Di Giorgio}, A. and {Guttridge}, P. and {James}, A. and {Kemp}, M. and {Martignac}, J. and {Spencer}, A. and {Walton}, D. and {Bl{\"u}mchen}, T. and {Bonoli}, C. and {Bortoletto}, F. and {Cerna}, C. and {Corcione}, L. and {Fabron}, C. and {Jahnke}, K. and {Ligori}, S. and {Madrid}, F. and {Martin}, L. and {Morgante}, G. and {Pamplona}, T. and {Prieto}, E. and {Riva}, M. and {Toledo}, R. and {Trifoglio}, M. and {Zerbi}, F. and {Abdalla}, F. and {Douspis}, M. and {Grenet}, C. and {Borgani}, S. and {Bouwens}, R. and {Courbin}, F. and {Delouis}, J. -M. and {Dubath}, P. and {Fontana}, A. and {Frailis}, M. and {Grazian}, A. and {Koppenh{\"o}fer}, J. and {Mansutti}, O. and {Melchior}, M. and {Mignoli}, M. and {Mohr}, J. and {Neissner}, C. and {Noddle}, K. and {Poncet}, M. and {Scodeggio}, M. and {Serrano}, S. and {Shane}, N. and {Starck}, J. -L. and {Surace}, C. and {Taylor}, A. and {Verdoes-Kleijn}, G. and {Vuerli}, C. and {Williams}, O.~R. and {Zacchei}, A. and {Altieri}, B. and {Escudero Sanz}, I. and {Kohley}, R. and {Oosterbroek}, T. and {Astier}, P. and {Bacon}, D. and {Bardelli}, S. and {Baugh}, C. and {Bellagamba}, F. and {Benoist}, C. and {Bianchi}, D. and {Biviano}, A. and {Branchini}, E. and {Carbone}, C. and {Cardone}, V. and {Clements}, D. and {Colombi}, S. and {Conselice}, C. and {Cresci}, G. and {Deacon}, N. and {Dunlop}, J. and {Fedeli}, C. and {Fontanot}, F. and {Franzetti}, P. and {Giocoli}, C. and {Garcia-Bellido}, J. and {Gow}, J. and {Heavens}, A. and {Hewett}, P. and {Heymans}, C. and {Holland}, A. and {Huang}, Z. and {Ilbert}, O. and {Joachimi}, B. and {Jennins}, E. and {Kerins}, E. and {Kiessling}, A. and {Kirk}, D. and {Kotak}, R. and {Krause}, O. and {Lahav}, O. and {van Leeuwen}, F. and {Lesgourgues}, J. and {Lombardi}, M. and {Magliocchetti}, M. and {Maguire}, K. and {Majerotto}, E. and {Maoli}, R. and {Marulli}, F. and {Maurogordato}, S. and {McCracken}, H. and {McLure}, R. and {Melchiorri}, A. and {Merson}, A. and {Moresco}, M. and {Nonino}, M. and {Norberg}, P. and {Peacock}, J. and {Pello}, R. and {Penny}, M. and {Pettorino}, V. and {Di Porto}, C. and {Pozzetti}, L. and {Quercellini}, C. and {Radovich}, M. and {Rassat}, A. and {Roche}, N. and {Ronayette}, S. and {Rossetti}, E. and {Sartoris}, B. and {Schneider}, P. and {Semboloni}, E. and {Serjeant}, S. and {Simpson}, F. and {Skordis}, C. and {Smadja}, G. and {Smartt}, S. and {Spano}, P. and {Spiro}, S. and {Sullivan}, M. and {Tilquin}, A. and {Trotta}, R. and {Verde}, L. and {Wang}, Y. and {Williger}, G. and {Zhao}, G. and {Zoubian}, J. and {Zucca}, E.},
        title = "{Euclid Definition Study Report}",
      journal = {arXiv e-prints},
         year = 2011,
        month = oct,
          eid = {arXiv:1110.3193},
        pages = {arXiv:1110.3193},
          doi = {10.48550/arXiv.1110.3193},
archivePrefix = {arXiv},
       eprint = {1110.3193},
 primaryClass = {astro-ph.CO},
       adsurl = {https://ui.adsabs.harvard.edu/abs/2011arXiv1110.3193L}
}

@article{EuclidNz,
  author = {{Euclid Collaboration}},
  title = {Euclid preparation: Mean redshift determination from galaxy redshift probabilities for cosmic shear tomography},
  journal = {Astronomy \& Astrophysics},
  year = {2020},
  volume = {642},
  pages = {A191}
}

@article{lsst2019,
  title={LSST: from science drivers to reference design and anticipated data products},
  author={Ivezi{\'c}, {\v{Z}}eljko and Kahn, Steven M and Tyson, J Anthony and Abel, Bob and Acosta, Emily and Allsman, Robyn and Alonso, David and AlSayyad, Yusra and Anderson, Scott F and Andrew, John and others},
  journal={The Astrophysical Journal},
  volume={873},
  number={2},
  pages={111},
  year={2019},
  publisher={The American Astronomical Society}
}

@inproceedings{RezendeMohamed2015,
  title={Variational inference with normalizing flows},
  author={Rezende, Danilo and Mohamed, Shakir},
  booktitle={International conference on machine learning},
  pages={1530--1538},
  year={2015},
  organization={PMLR}
}

@article{Photoz2019,
  author = {Salvato, M. and others},
  title = {The many flavours of photometric redshifts},
  journal = {Nature Astronomy},
  year = {2019},
  volume = {3},
  pages = {212--222}
}

@article{Photoz2022,
  title={Photometric redshifts for next-generation surveys},
  author={Newman, Jeffrey A and Gruen, Daniel},
  journal={Annual Review of Astronomy and Astrophysics},
  volume={60},
  number={1},
  pages={363--414},
  year={2022},
  publisher={Annual Reviews}
}

@article{RF2001,
  author = {Breiman, L.},
  title = {Random Forests},
  journal = {Machine Learning},
  year = {2001},
  volume = {45},
  pages = {5--32}
}

@article{QRF2006,
  author = {Meinshausen, N.},
  title = {Quantile Regression Forests},
  journal = {Journal of Machine Learning Research},
  year = {2006},
  volume = {7},
  pages = {983--999}
}

@article{NF2019,
  author = {Durkan, C. and Bekasov, A. and Murray, I. and Papamakarios, G.},
  title = {Neural Spline Flows},
  journal = {NeurIPS},
  year = {2019}
}

@ARTICLE{ren2026NF,
       author = {{Ren}, Yiming and {Chan}, Kwan Chuen and {Zhang}, Le and {Li}, Yin and others},
        title = "{Photo-z Estimation with Normalizing Flow}",
      journal = {\apj},
         year = 2026,
        month = jan,
       volume = {997},
       number = {1},
          eid = {65},
        pages = {65},
          doi = {10.3847/1538-4357/ae201d},
archivePrefix = {arXiv},
       eprint = {2510.10032},
 primaryClass = {astro-ph.IM},
       adsurl = {https://ui.adsabs.harvard.edu/abs/2026ApJ...997...65R}
}

@article{Dawid1984PIT,
  author = {Dawid, A. P.},
  title = {Statistical Theory: The Prequential Approach},
  journal = {Journal of the Royal Statistical Society A},
  year = {1984},
  volume = {147},
  pages = {278--292}
}

@article{Wittman2016PIT,
  author = {Wittman, D. and others},
  title = {Using nearest neighbors to estimate redshift distributions},
  journal = {MNRAS},
  year = {2016},
  volume = {457},
  pages = {4005--4015}
}

@article{rau2015accurate,
  title   = {Accurate photometric redshift probability density estimation--method comparison and application},
  author  = {Rau, MM and Seitz, S and Brimioulle, F and Frank, E},
  journal = {Monthly Notices of the Royal Astronomical Society},
  volume  = {452},
  pages   = {3710--3724},
  year    = {2015},
  doi     = {10.1093/mnras/stv1567},
  url     = {https://academic.oup.com/mnras/article-pdf/452/4/3710/18241297/stv1567.pdf}
}

@inproceedings{guo2017calibration,
  title     = {On Calibration of Modern Neural Networks},
  author    = {Guo, Chuan and Pleiss, Geoff and Sun, Yu and Weinberger, Kilian Q.},
  booktitle = {Proceedings of the 34th International Conference on Machine Learning (ICML)},
  year      = {2017},
  volume    = {70},
  pages     = {1321--1330},
  url       = {http://proceedings.mlr.press/v70/guo17a.html}
}

@inproceedings{niculescu2005predicting,
  title     = {Predicting Good Probabilities with Supervised Learning},
  author    = {Niculescu-Mizil, Alex and Caruana, Rich},
  booktitle = {Proceedings of the 22nd International Conference on Machine Learning (ICML)},
  year      = {2005},
  pages     = {625--632},
  url       = {https://dl.acm.org/doi/10.1145/1102351.1102430}
}

@inproceedings{simonyan2014deep,
  title     = {Deep Inside Convolutional Networks: Visualising Image Classification Models and Saliency Maps},
  author    = {Simonyan, Karen and Vedaldi, Andrea and Zisserman, Andrew},
  booktitle = {Workshop at International Conference on Learning Representations (ICLR)},
  year      = {2014},
  url       = {https://arxiv.org/abs/1312.6034}
}

@book{Villani2009wat1,
  author = {Villani, C.},
  title = {Optimal Transport: Old and New},
  publisher = {Springer},
  year = {2009}
}

@article{Lin1991JW,
  author = {Lin, J.},
  title = {Divergence measures based on the Shannon entropy},
  journal = {IEEE Transactions on Information Theory},
  year = {1991},
  volume = {37},
  pages = {145--151}
}

@article{Kullback1951overlap,
  author = {Kullback, S. and Leibler, R. A.},
  title = {On Information and Sufficiency},
  journal = {Annals of Mathematical Statistics},
  year = {1951},
  volume = {22},
  pages = {79--86}
}

@article{Schlegel1998Dust,
  author  = {Schlegel, David J. and Finkbeiner, Douglas P. and Davis, Marc},
  title   = {Maps of Dust Infrared Emission for Use in Estimation of Reddening and Cosmic Microwave Background Radiation Foregrounds},
  journal = {The Astrophysical Journal},
  year    = {1998},
  volume  = {500},
  number  = {2},
  pages   = {525--553},
  doi     = {10.1086/305772},
  url     = {https://iopscience.iop.org/article/10.1086/305772}
}

@article{Schlafly2011extinction,
  author = {Schlafly, E. F. and Finkbeiner, D. P.},
  title = {Measuring Reddening with SDSS},
  journal = {ApJ},
  year = {2011},
  volume = {737},
  pages = {103}
}

@ARTICLE{2026tian,
       author = {{Tian}, Da-Chuan and {Wen}, Zhong-Lue and {Xia}, Jun-Qing},
        title = "{Photometric Redshift PDFs via Neural Network Classification for DESI Legacy Imaging Surveys and Pan-STARRS}",
      journal = {arXiv e-prints},
         year = 2026,
        month = feb,
          eid = {arXiv:2602.01548},
        pages = {arXiv:2602.01548},
          doi = {10.48550/arXiv.2602.01548},
archivePrefix = {arXiv},
       eprint = {2602.01548},
 primaryClass = {astro-ph.GA},
       adsurl = {https://ui.adsabs.harvard.edu/abs/2026arXiv260201548T}
}

@article{DIsanto2018,
  author  = {{D'Isanto}, A. and {Polsterer}, K.~L.},
  title   = {Photometric redshift estimation via deep learning: Generalized and pre-classification-less, image-based, fully probabilistic redshifts},
  journal = {Astronomy \& Astrophysics},
  volume  = {609},
  pages   = {A111},
  year    = {2018},
  doi     = {10.1051/0004-6361/201731326},
  eprint  = {1706.02467},
  archivePrefix = {arXiv},
  primaryClass  = {astro-ph.IM}
}

@inproceedings{kuleshov2018accurate,
    author    = {Kuleshov, Volodymyr and Fenner, Nathan and Ermon, Stefano},
    title     = {Accurate Uncertainties for Deep Learning Using Calibrated Regression},
    booktitle = {Proceedings of the 35th International Conference on Machine Learning},
    series    = {Proceedings of Machine Learning Research},
    volume    = {80},
    pages     = {2796--2804},
    year      = {2018},
    publisher = {PMLR}
}

@ARTICLE{Crenshaw_etal2024,
       author = {{Crenshaw}, John Franklin and {Kalmbach}, J. Bryce and {Gagliano}, Alexander and {Yan}, Ziang and {Connolly}, Andrew J. and {Malz}, Alex I. and {Schmidt}, Samuel J. and {The LSST Dark Energy Science Collaboration}},
        title = "{Probabilistic Forward Modeling of Galaxy Catalogs with Normalizing Flows}",
      journal = {\aj},
         year = 2024,
        month = aug,
       volume = {168},
       number = {2},
          eid = {80},
        pages = {80},
          doi = {10.3847/1538-3881/ad54bf},
archivePrefix = {arXiv},
       eprint = {2405.04740},
 primaryClass = {astro-ph.IM},
       adsurl = {https://ui.adsabs.harvard.edu/abs/2024AJ....168...80C}
}

@ARTICLE{Zheng_etal2024,
       author = {{Zheng}, Weilun and {Chan}, Kwan Chuen and {Xu}, Haojie and {Zhang}, Le and {Song}, Ruiyu},
        title = "{Optimizing redshift distribution inference through joint self-calibration and clustering-redshift synergy}",
      journal = {\aap},
         year = 2024,
        month = dec,
       volume = {692},
          eid = {A186},
        pages = {A186},
          doi = {10.1051/0004-6361/202452334},
archivePrefix = {arXiv},
       eprint = {2409.12009},
 primaryClass = {astro-ph.CO},
       adsurl = {https://ui.adsabs.harvard.edu/abs/2024A&A...692A.186Z}
}

@ARTICLE{XU2023,
       author = {{Xu}, Haojie and {Zhang}, Pengjie and {Peng}, Hui and {Yu}, Yu and {Zhang}, Le and {Yao}, Ji and {Qin}, Jian and {Sun}, Zeyang and {He}, Min and {Yang}, Xiaohu},
        title = "{Using angular two-point correlations to self-calibrate the photometric redshift distributions of DECaLS DR9}",
      journal = {MNRAS},
     fjournal = {Monthly Notices of the Royal Astronomial Society},
         year = 2023,
        month = mar,
       volume = {520},
       number = {1},
        pages = {161-179},
          doi = {10.1093/mnras/stad136},
archivePrefix = {arXiv},
       eprint = {2209.03967},
 primaryClass = {astro-ph.CO},
       adsurl = {https://ui.adsabs.harvard.edu/abs/2023MNRAS.520..161X}
}

@ARTICLE{ Schneider_etal2006,
       author = {{Schneider}, Michael and {Knox}, Lloyd and {Zhan}, Hu and {Connolly}, Andrew},
        title = "{Using Galaxy Two-Point Correlation Functions to Determine the Redshift Distributions of Galaxies Binned by Photometric Redshift}",
      journal = {\apj},
         year = 2006,
        month = nov,
       volume = {651},
       number = {1},
        pages = {14-23},
          doi = {10.1086/507675},
archivePrefix = {arXiv},
       eprint = {astro-ph/0606098},
 primaryClass = {astro-ph},
       adsurl = {https://ui.adsabs.harvard.edu/abs/2006ApJ...651...14S}
}

@ARTICLE{Zhang_etal2017,
       author = {{Zhang}, Le and {Yu}, Yu and {Zhang}, Pengjie},
        title = "{Non-negative Matrix Factorization for Self-calibration of Photometric Redshift Scatter in Weak-lensing Surveys}",
      journal = {\apj},
         year = 2017,
        month = oct,
       volume = {848},
       number = {1},
          eid = {44},
        pages = {44},
          doi = {10.3847/1538-4357/aa8c72},
archivePrefix = {arXiv},
       eprint = {1612.04042},
 primaryClass = {astro-ph.CO},
       adsurl = {https://ui.adsabs.harvard.edu/abs/2017ApJ...848...44Z}
}

@ARTICLE{Zhang_etal2010,
       author = {{Zhang}, Pengjie and {Pen}, Ue-Li and {Bernstein}, Gary},
        title = "{Self-calibration of photometric redshift scatter in weak-lensing surveys}",
      journal = {\mnras},
         year = 2010,
        month = jun,
       volume = {405},
       number = {1},
        pages = {359-374},
          doi = {10.1111/j.1365-2966.2010.16445.x},
archivePrefix = {arXiv},
       eprint = {0910.4181},
 primaryClass = {astro-ph.CO},
       adsurl = {https://ui.adsabs.harvard.edu/abs/2010MNRAS.405..359Z}
}

@ARTICLE{Peng_etal2022,
       author = {{Peng}, Hui and {Xu}, Haojie and {Zhang}, Le and {Chen}, Zhao and {Yu}, Yu},
        title = "{Self-calibration of photometric redshift scatter from DECaLS DR8 power spectrum and validation with simulated catalogues}",
      journal = {\mnras},
         year = 2022,
        month = nov,
       volume = {516},
       number = {4},
        pages = {6210-6224},
          doi = {10.1093/mnras/stac2713},
archivePrefix = {arXiv},
       eprint = {2203.14609},
 primaryClass = {astro-ph.CO},
       adsurl = {https://ui.adsabs.harvard.edu/abs/2022MNRAS.516.6210P}
}

@ARTICLE{Newman_2008,
       author = {{Newman}, Jeffrey A.},
        title = "{Calibrating Redshift Distributions beyond Spectroscopic Limits with Cross-Correlations}",
      journal = {\apj},
         year = 2008,
        month = sep,
       volume = {684},
       number = {1},
        pages = {88-101},
          doi = {10.1086/589982},
archivePrefix = {arXiv},
       eprint = {0805.1409},
 primaryClass = {astro-ph},
       adsurl = {https://ui.adsabs.harvard.edu/abs/2008ApJ...684...88N}
}

@ARTICLE{ Schmidt_etal2013,
       author = {{Schmidt}, Samuel J. and {M{\'e}nard}, Brice and {Scranton}, Ryan and {Morrison}, Christopher and {McBride}, Cameron K.},
        title = "{Recovering redshift distributions with cross-correlations: pushing the boundaries}",
      journal = {\mnras},
         year = 2013,
        month = jun,
       volume = {431},
       number = {4},
        pages = {3307-3318},
          doi = {10.1093/mnras/stt410},
archivePrefix = {arXiv},
       eprint = {1303.0292},
 primaryClass = {astro-ph.CO},
       adsurl = {https://ui.adsabs.harvard.edu/abs/2013MNRAS.431.3307S}
}

@ARTICLE{ McQuinnWhite_2013,
       author = {{McQuinn}, Matthew and {White}, Martin},
        title = "{On using angular cross-correlations to determine source redshift distributions}",
      journal = {\mnras},
         year = 2013,
        month = aug,
       volume = {433},
       number = {4},
        pages = {2857-2883},
          doi = {10.1093/mnras/stt914},
archivePrefix = {arXiv},
       eprint = {1302.0857},
 primaryClass = {astro-ph.CO},
       adsurl = {https://ui.adsabs.harvard.edu/abs/2013MNRAS.433.2857M}
}

@ARTICLE{ Menard_etal2013,
       author = {{M{\'e}nard}, Brice and {Scranton}, Ryan and {Schmidt}, Samuel and {Morrison}, Chris and {Jeong}, Donghui and {Budavari}, Tamas and {Rahman}, Mubdi},
        title = "{Clustering-based redshift estimation: method and application to data}",
      journal = {arXiv e-prints},
         year = 2013,
        month = mar,
          eid = {arXiv:1303.4722},
        pages = {arXiv:1303.4722},
          doi = {10.48550/arXiv.1303.4722},
archivePrefix = {arXiv},
       eprint = {1303.4722},
 primaryClass = {astro-ph.CO},
       adsurl = {https://ui.adsabs.harvard.edu/abs/2013arXiv1303.4722M}
}

@ARTICLE{Wright_etal2020,
       author = {{Wright}, Angus H. and {Hildebrandt}, Hendrik and {van den Busch}, Jan Luca and {Heymans}, Catherine},
        title = "{Photometric redshift calibration with self-organising maps}",
      journal = {\aap},
         year = 2020,
        month = may,
       volume = {637},
          eid = {A100},
        pages = {A100},
          doi = {10.1051/0004-6361/201936782},
archivePrefix = {arXiv},
       eprint = {1909.09632},
 primaryClass = {astro-ph.CO},
       adsurl = {https://ui.adsabs.harvard.edu/abs/2020A&A...637A.100W}
}

@ARTICLE{Buchs_etal2019,
       author = {{Buchs}, R. and {Davis}, C. and {Gruen}, D. and {DeRose}, J. and {Alarcon}, A. and {Bernstein}, G.~M. and {S{\'a}nchez}, C. and {Myles}, J. and {Roodman}, A. and {Allen}, S. and {Amon}, A. and {Choi}, A. and {Masters}, D.~C. and {Miquel}, R. and {Troxel}, M.~A. and {Wechsler}, R.~H. and {Abbott}, T.~M.~C. and {Annis}, J. and {Avila}, S. and {Bechtol}, K. and {Bridle}, S.~L. and {Brooks}, D. and {Buckley-Geer}, E. and {Burke}, D.~L. and {Carnero Rosell}, A. and {Carrasco Kind}, M. and {Carretero}, J. and {Castander}, F.~J. and {Cawthon}, R. and {D'Andrea}, C.~B. and {da Costa}, L.~N. and {De Vicente}, J. and {Desai}, S. and {Diehl}, H.~T. and {Doel}, P. and {Drlica-Wagner}, A. and {Eifler}, T.~F. and {Evrard}, A.~E. and {Flaugher}, B. and {Fosalba}, P. and {Frieman}, J. and {Garc{\'\i}a-Bellido}, J. and {Gaztanaga}, E. and {Gruendl}, R.~A. and {Gschwend}, J. and {Gutierrez}, G. and {Hartley}, W.~G. and {Hollowood}, D.~L. and {Honscheid}, K. and {James}, D.~J. and {Kuehn}, K. and {Kuropatkin}, N. and {Lima}, M. and {Lin}, H. and {Maia}, M.~A.~G. and {March}, M. and {Marshall}, J.~L. and {Melchior}, P. and {Menanteau}, F. and {Ogando}, R.~L.~C. and {Plazas}, A.~A. and {Rykoff}, E.~S. and {Sanchez}, E. and {Scarpine}, V. and {Serrano}, S. and {Sevilla-Noarbe}, I. and {Smith}, M. and {Soares-Santos}, M. and {Sobreira}, F. and {Suchyta}, E. and {Swanson}, M.~E.~C. and {Tarle}, G. and {Thomas}, D. and {Vikram}, V. and {DES Collaboration}},
        title = "{Phenotypic redshifts with self-organizing maps: A novel method to characterize redshift distributions of source galaxies for weak lensing}",
      journal = {\mnras},
         year = 2019,
        month = oct,
       volume = {489},
       number = {1},
        pages = {820-841},
          doi = {10.1093/mnras/stz2162},
archivePrefix = {arXiv},
       eprint = {1901.05005},
 primaryClass = {astro-ph.CO},
       adsurl = {https://ui.adsabs.harvard.edu/abs/2019MNRAS.489..820B}
}

@ARTICLE{Masters_etal2015,
       author = {{Masters}, Daniel and {Capak}, Peter and {Stern}, Daniel and {Ilbert}, Olivier and {Salvato}, Mara and {Schmidt}, Samuel and {Longo}, Giuseppe and {Rhodes}, Jason and {Paltani}, Stephane and {Mobasher}, Bahram and {Hoekstra}, Henk and {Hildebrandt}, Hendrik and {Coupon}, Jean and {Steinhardt}, Charles and {Speagle}, Josh and {Faisst}, Andreas and {Kalinich}, Adam and {Brodwin}, Mark and {Brescia}, Massimo and {Cavuoti}, Stefano},
        title = "{Mapping the Galaxy Color-Redshift Relation: Optimal Photometric Redshift Calibration Strategies for Cosmology Surveys}",
      journal = {\apj},
         year = 2015,
        month = nov,
       volume = {813},
       number = {1},
          eid = {53},
        pages = {53},
          doi = {10.1088/0004-637X/813/1/53},
archivePrefix = {arXiv},
       eprint = {1509.03318},
 primaryClass = {astro-ph.CO},
       adsurl = {https://ui.adsabs.harvard.edu/abs/2015ApJ...813...53M}
}

@ARTICLE{KindBrunner_2014,
       author = {{Carrasco Kind}, Matias and {Brunner}, Robert J.},
        title = "{SOMz: photometric redshift PDFs with self-organizing maps and random atlas}",
      journal = {\mnras},
         year = 2014,
        month = mar,
       volume = {438},
       number = {4},
        pages = {3409-3421},
          doi = {10.1093/mnras/stt2456},
archivePrefix = {arXiv},
       eprint = {1312.5753},
 primaryClass = {astro-ph.IM},
       adsurl = {https://ui.adsabs.harvard.edu/abs/2014MNRAS.438.3409C}
}

@ARTICLE{CarrascoKind_Brunner2013,
       author = {{Carrasco Kind}, Matias and {Brunner}, Robert J.},
        title = "{TPZ: photometric redshift PDFs and ancillary information by using prediction trees and random forests}",
      journal = {\mnras},
         year = 2013,
        month = jun,
       volume = {432},
       number = {2},
        pages = {1483-1501},
          doi = {10.1093/mnras/stt574},
archivePrefix = {arXiv},
       eprint = {1303.7269},
 primaryClass = {astro-ph.CO},
       adsurl = {https://ui.adsabs.harvard.edu/abs/2013MNRAS.432.1483C}
}

@ARTICLE{Almosallam_etal2016,
       author = {{Almosallam}, Ibrahim A. and {Jarvis}, Matt J. and {Roberts}, Stephen J.},
        title = "{GPZ: non-stationary sparse Gaussian processes for heteroscedastic uncertainty estimation in photometric redshifts}",
      journal = {\mnras},
         year = 2016,
        month = oct,
       volume = {462},
       number = {1},
        pages = {726-739},
          doi = {10.1093/mnras/stw1618},
archivePrefix = {arXiv},
       eprint = {1604.03593},
 primaryClass = {astro-ph.IM},
       adsurl = {https://ui.adsabs.harvard.edu/abs/2016MNRAS.462..726A}
}

@ARTICLE{Schuldt_etal2021,
       author = {{Schuldt}, S. and {Suyu}, S.~H. and {Ca{\~n}ameras}, R. and {Taubenberger}, S. and {Meinhardt}, T. and {Leal-Taix{\'e}}, L. and {Hsieh}, B.~C.},
        title = "{Photometric redshift estimation with a convolutional neural network: NetZ}",
      journal = {\aap},
         year = 2021,
        month = jul,
       volume = {651},
          eid = {A55},
        pages = {A55},
          doi = {10.1051/0004-6361/202039945},
archivePrefix = {arXiv},
       eprint = {2011.12312},
 primaryClass = {astro-ph.GA},
       adsurl = {https://ui.adsabs.harvard.edu/abs/2021A&A...651A..55S}
}

@ARTICLE{Gerdes_etal2010,
       author = {{Gerdes}, David W. and {Sypniewski}, Adam J. and {McKay}, Timothy A. and {Hao}, Jiangang and {Weis}, Matthew R. and {Wechsler}, Risa H. and {Busha}, Michael T.},
        title = "{ArborZ: Photometric Redshifts Using Boosted Decision Trees}",
      journal = {\apj},
         year = 2010,
        month = jun,
       volume = {715},
       number = {2},
        pages = {823-832},
          doi = {10.1088/0004-637X/715/2/823},
archivePrefix = {arXiv},
       eprint = {0908.4085},
 primaryClass = {astro-ph.CO},
       adsurl = {https://ui.adsabs.harvard.edu/abs/2010ApJ...715..823G}
}

@ARTICLE{Hoyle_2016,
       author = {{Hoyle}, B.},
        title = "{Measuring photometric redshifts using galaxy images and Deep Neural Networks}",
      journal = {Astronomy and Computing},
         year = 2016,
        month = jul,
       volume = {16},
        pages = {34-40},
          doi = {10.1016/j.ascom.2016.03.006},
archivePrefix = {arXiv},
       eprint = {1504.07255},
 primaryClass = {astro-ph.IM},
       adsurl = {https://ui.adsabs.harvard.edu/abs/2016A&C....16...34H}
}

@ARTICLE{LiNapolitano_etal2022,
       author = {{Li}, R. and {Napolitano}, N.~R. and {Roy}, N. and {Tortora}, C. and {La Barbera}, F. and {Sonnenfeld}, A. and {Qiu}, C. and {Liu}, S.},
        title = "{Galaxy Light Profile Convolutional Neural Networks (GaLNets). I. Fast and Accurate Structural Parameters for Billion-galaxy Samples}",
      journal = {\apj},
         year = 2022,
        month = apr,
       volume = {929},
       number = {2},
          eid = {152},
        pages = {152},
          doi = {10.3847/1538-4357/ac5ea0},
archivePrefix = {arXiv},
       eprint = {2111.05434},
 primaryClass = {astro-ph.GA},
       adsurl = {https://ui.adsabs.harvard.edu/abs/2022ApJ...929..152L}
}

@ARTICLE{ZhouNewman_etal2021,
       author = {{Zhou}, Rongpu and {Newman}, Jeffrey A. and {Mao}, Yao-Yuan and {Meisner}, Aaron and {Moustakas}, John and {Myers}, Adam D. and {Prakash}, Abhishek and {Zentner}, Andrew R. and {Brooks}, David and {Duan}, Yutong and {Landriau}, Martin and {Levi}, Michael E. and {Prada}, Francisco and {Tarle}, Gregory},
        title = "{The clustering of DESI-like luminous red galaxies using photometric redshifts}",
      journal = {\mnras},
         year = 2021,
        month = mar,
       volume = {501},
       number = {3},
        pages = {3309-3331},
          doi = {10.1093/mnras/staa3764},
archivePrefix = {arXiv},
       eprint = {2001.06018},
 primaryClass = {astro-ph.CO},
       adsurl = {https://ui.adsabs.harvard.edu/abs/2021MNRAS.501.3309Z}
}

@ARTICLE{Carliles_etal2010,
       author = {{Carliles}, Samuel and {Budav{\'a}ri}, Tam{\'a}s and {Heinis}, S{\'e}bastien and {Priebe}, Carey and {Szalay}, Alexander S.},
        title = "{Random Forests for Photometric Redshifts}",
      journal = {\apj},
         year = 2010,
        month = mar,
       volume = {712},
       number = {1},
        pages = {511-515},
          doi = {10.1088/0004-637X/712/1/511},
       adsurl = {https://ui.adsabs.harvard.edu/abs/2010ApJ...712..511C}
}

@ARTICLE{CollisterLahav_2004,
       author = {{Collister}, Adrian A. and {Lahav}, Ofer},
        title = "{ANNz: Estimating Photometric Redshifts Using Artificial Neural Networks}",
      journal = {\pasp},
         year = 2004,
        month = apr,
       volume = {116},
       number = {818},
        pages = {345-351},
          doi = {10.1086/383254},
archivePrefix = {arXiv},
       eprint = {astro-ph/0311058},
 primaryClass = {astro-ph},
       adsurl = {https://ui.adsabs.harvard.edu/abs/2004PASP..116..345C}
}

@article{Sadeh2016,
    author  = {Sadeh, I. and Abdalla, F. B. and Lahav, O.},
    title   = {{ANNz2}: Photometric Redshift and Probability Distribution
               Function Estimation using Machine Learning},
    journal = {PASP},
    year    = {2016},
    volume  = {128},
    number  = {968},
    pages   = {104502},
    doi     = {10.1088/1538-3873/128/968/104502},
    eprint  = {1507.00490},
    archivePrefix = {arXiv}
}

@article{DeVicente:2015kyp,
    author = "De Vicente, Juan and S\'anchez, Eusebio and Sevilla-Noarbe, Ignacio",
    title = "{DNF \textendash{} Galaxy photometric redshift by Directional Neighbourhood Fitting}",
    eprint = "1511.07623",
    archivePrefix = "arXiv",
    primaryClass = "astro-ph.CO",
    doi = "10.1093/mnras/stw857",
    journal = "Mon. Not. Roy. Astron. Soc.",
    volume = "459",
    number = "3",
    pages = "3078--3088",
    year = "2016"
}

@ARTICLE{CSST_2026,
       author = {{CSST Collaboration} and {Gong}, Yan and {Miao}, Haitao and {Zhan}, Hu and {Li}, Zhao-Yu and {Shangguan}, Jinyi and {Li}, Haining and {Liu}, Chao and {Chen}, Xuefei and {Yuan}, Haibo and et al.},
        title = "{Introduction to the Chinese Space Station Survey Telescope (CSST)}",
      journal = {Science China Physics, Mechanics, and Astronomy},
         year = 2026,
        month = jan,
       volume = {69},
       number = {3},
          eid = {239501},
        pages = {239501},
          doi = {10.1007/s11433-025-2809-0},
archivePrefix = {arXiv},
       eprint = {2507.04618},
 primaryClass = {astro-ph.IM},
       adsurl = {https://ui.adsabs.harvard.edu/abs/2026SCPMA..6939501C}
}

@article{Tabak:2013cnz,
    author = "Tabak, E. G. and Turner, Cristina V.",
    title = "{A Family of Nonparametric Density Estimation Algorithms}",
    doi = "10.1002/cpa.21423",
    journal = "Commun. Pure Appl. Math.",
    volume = "66",
    number = "2",
    pages = "145--164",
    year = "2013"
}

@ARTICLE{Dey_etal2021,
       author = {{Dey}, Biprateep and {Newman}, Jeffrey A. and {Andrews}, Brett H. and {Izbicki}, Rafael and {Lee}, Ann B. and {Zhao}, David and {Rau}, Markus Michael and {Malz}, Alex I.},
        title = "{Re-calibrating Photometric Redshift Probability Distributions Using Feature-space Regression}",
      journal = {arXiv e-prints},
         year = 2021,
        month = oct,
          eid = {arXiv:2110.15209},
        pages = {arXiv:2110.15209},
          doi = {10.48550/arXiv.2110.15209},
archivePrefix = {arXiv},
       eprint = {2110.15209},
 primaryClass = {astro-ph.IM},
       adsurl = {https://ui.adsabs.harvard.edu/abs/2021arXiv211015209D}
}

@ARTICLE{Sun_etal2023,
       author = {{Sun}, Zechang and {Speagle}, Joshua S. and {Huang}, Song and {Ting}, Yuan-Sen and {Cai}, Zheng},
        title = "{Zephyr : Stitching Heterogeneous Training Data with Normalizing Flows for Photometric Redshift Inference}",
      journal = {arXiv e-prints},
         year = 2023,
        month = oct,
          eid = {arXiv:2310.20125},
        pages = {arXiv:2310.20125},
          doi = {10.48550/arXiv.2310.20125},
archivePrefix = {arXiv},
       eprint = {2310.20125},
 primaryClass = {astro-ph.IM},
       adsurl = {https://ui.adsabs.harvard.edu/abs/2023arXiv231020125S}
}

@ARTICLE{Zhou_etal2022,
       author = {{Zhou}, Xingchen and {Gong}, Yan and {Meng}, Xian-Min and {Cao}, Ye and {Chen}, Xuelei and {Chen}, Zhu and {Du}, Wei and {Fu}, Liping and {Luo}, Zhijian},
        title = "{Extracting photometric redshift from galaxy flux and image data using neural networks in the CSST survey}",
      journal = {\mnras},
         year = 2022,
        month = may,
       volume = {512},
       number = {3},
        pages = {4593-4603},
          doi = {10.1093/mnras/stac786},
archivePrefix = {arXiv},
       eprint = {2112.08690},
 primaryClass = {astro-ph.CO},
       adsurl = {https://ui.adsabs.harvard.edu/abs/2022MNRAS.512.4593Z}
}

@ARTICLE{Lu_etal2024,
       author = {{Lu}, Junhao and {Luo}, Zhijian and {Chen}, Zhu and {Fu}, Liping and {Du}, Wei and {Gong}, Yan and {Li}, Yicheng and {Meng}, Xian-Min and {Tang}, Zhirui and {Zhang}, Shaohua and {Shu}, Chenggang and {Zhou}, Xingchen and {Fan}, Zuhui},
        title = "{Estimating photometric redshift from mock flux for CSST survey by using weighted Random Forest}",
      journal = {\mnras},
         year = 2024,
        month = feb,
       volume = {527},
       number = {4},
        pages = {12140-12153},
          doi = {10.1093/mnras/stad3976},
archivePrefix = {arXiv},
       eprint = {2312.12949},
 primaryClass = {astro-ph.CO},
       adsurl = {https://ui.adsabs.harvard.edu/abs/2024MNRAS.52712140L}
}

@article{Euclid2021Photoz,
    author  = {{Euclid Collaboration}},
    title   = {Euclid preparation. XI. Mean redshift determination from
               galaxy redshift probabilities for cosmic shear tomography},
    journal = {A\&A},
    year    = {2021},
    volume  = {647},
    pages   = {A117},
    doi     = {10.1051/0004-6361/202040237},
    eprint  = {2101.02228},
    archivePrefix = {arXiv}
}

@inproceedings{Dinh2014NICE,
  author    = {Dinh, Laurent and Krueger, David and Bengio, Yoshua},
  title     = {{NICE}: Non-linear Independent Components Estimation},
  booktitle = {International Conference on Learning Representations (ICLR), Workshop Track},
  year      = {2015},
  eprint    = {1410.8516},
  archivePrefix = {arXiv},
  primaryClass  = {cs.LG},
  url       = {https://arxiv.org/abs/1410.8516}
}

@article{Lange2026,
    author = {Lange, J. U. and Blanco, D. and Leauthaud, A.
              and Wright, A. H. and Fisher, A.
              and Ratajczak, J. and others},
    title = {A Unified Photometric Redshift Calibration
             for Weak Lensing Surveys Using the
             Dark Energy Spectroscopic Instrument},
    journal = {The Astrophysical Journal},
    year = {2026},
    volume = {998},
    number = {2},
    pages = {263},
    doi = {10.3847/1538-4357/ae3165},
    eprint = {2510.25419},
    archivePrefix = {arXiv}
}

@article{Blanco2026,
    author = {Blanco, D. and Leauthaud, A. and Lange, J. U.
              and Wright, A. H. and Hildebrandt, H.
              and Heydenreich, S. and others},
    title = {The Power of DESI for Photometric Redshift
             Calibration: A Case Study with KiDS-1000},
    journal = {The Open Journal of Astrophysics},
    year = {2026},
    volume = {9},
    doi = {10.33232/001c.163965},
    eprint = {2512.15964},
    archivePrefix = {arXiv}
}

@article{ChoppindeJanvry2026,
  author = {{Choppin de Janvry}, J. and
            {Gontcho A Gontcho}, S. and
            Seljak, U. and
            {Baleato Lizancos}, A. and
            Chaussidon, E. and others},
  title = {Full calibration of the tomographic
           redshift distribution from the HSC
           PDR3 Shape Catalog with DESI},
  journal = {Journal of Cosmology and
             Astroparticle Physics},
  volume = {2026},
  number = {05},
  pages = {073},
  year = {2026},
  doi = {10.1088/1475-7516/2026/05/073},
  eprint = {2511.18133},
  archivePrefix = {arXiv}
}

@article{hahn_desi_2023,
  title = {{DESI} {Bright} {Galaxy} {Survey}: {Final} {Target} {Selection}, {Design}, and {Validation}},
  volume = {165},
  issn = {0004-6256, 1538-3881},
  shorttitle = {{DESI} {Bright} {Galaxy} {Survey}},
  url = {http://arxiv.org/abs/2208.08512},
  doi = {10.3847/1538-3881/accff8},
  number = {6},
  urldate = {2026-07-08},
  journal = {The Astronomical Journal},
  author = {Hahn, ChangHoon and Wilson, Michael J. and Ruiz-Macias, Omar and Cole, Shaun and Weinberg, David H. and Moustakas, John and Kremin, Anthony and Tinker, Jeremy L. and Smith, Alex and Wechsler, Risa H. and Ahlen, Steven and Alam, Shadab and Bailey, Stephen and Brooks, David and Cooper, Andrew P. and Davis, Tamara M. and Dawson, Kyle and Dey, Arjun and Dey, Biprateep and Eftekharzadeh, Sarah and Eisenstein, Daniel J. and Fanning, Kevin and Forero-Romero, Jaime E. and Frenk, Carlos S. and Gaztañaga, Enrique and Gontcho, Satya Gontcho A. and Guy, Julien and Honscheid, Klaus and Ishak, Mustapha and Juneau, Stéphanie and Kehoe, Robert and Kisner, Theodore and Lan, Ting-Wen and Landriau, Martin and Guillou, Laurent Le and Levi, Michael E. and Magneville, Christophe and Martini, Paul and Meisner, Aaron and Myers, Adam D. and Nie, Jundan and Norberg, Peder and Palanque-Delabrouille, Nathalie and Percival, Will J. and Poppett, Claire and Prada, Francisco and Raichoor, Anand and Ross, Ashley J. and Safonova, Sasha and Saulder, Christoph and Schlafly, Eddie and Schlegel, David and Sierra-Porta, David and Tarle, Gregory and Weaver, Benjamin A. and Yèche, Christophe and Zarrouk, Pauline and Zhou, Rongpu and Zhou, Zhimin and Zou, Hu},
  month = jun,
  year = {2023},
  note = {arXiv:2208.08512 [astro-ph.CO]},
  pages = {253},
}

@article{zhou_target_2023,
  title = {Target {Selection} and {Validation} of {DESI} {Luminous} {Red} {Galaxies}},
  volume = {165},
  issn = {0004-6256, 1538-3881},
  url = {http://arxiv.org/abs/2208.08515},
  doi = {10.3847/1538-3881/aca5fb},
  language = {en},
  number = {2},
  urldate = {2026-07-08},
  journal = {The Astronomical Journal},
  author = {Zhou, Rongpu and Dey, Biprateep and Newman, Jeffrey A. and Eisenstein, Daniel J. and Dawson, K. and Bailey, S. and Berti, A. and Guy, J. and Lan, Ting-Wen and Zou, H. and Aguilar, J. and Ahlen, S. and Alam, Shadab and Brooks, D. and Macorra, A. de la and Dey, A. and Dhungana, G. and Fanning, K. and Font-Ribera, A. and Gontcho, S. Gontcho A. and Honscheid, K. and Ishak, Mustapha and Kisner, T. and Kovács, A. and Kremin, A. and Landriau, M. and Levi, Michael E. and Magneville, C. and Manera, Marc and Martini, P. and Meisner, Aaron M. and Miquel, R. and Moustakas, J. and Myers, Adam D. and Nie, Jundan and Palanque-Delabrouille, N. and Percival, W. J. and Poppett, C. and Prada, F. and Raichoor, A. and Ross, A. J. and Schlafly, E. and Schlegel, D. and Schubnell, M. and Tarlé, Gregory and Weaver, B. A. and Wechsler, R. H. and Yèche, Christophe and Zhou, Zhimin},
  month = feb,
  year = {2023},
  note = {arXiv:2208.08515 [astro-ph.CO]},
  pages = {58},
}

@misc{chaussidon_target_2022,
  title = {Target {Selection} and {Validation} of {DESI} {Quasars}},
  url = {https://arxiv.org/abs/2208.08511v2},
  doi = {10.3847/1538-4357/acb3c2},
  language = {en},
  urldate = {2026-08-03},
  journal = {arXiv.org},
  author = {Chaussidon, Edmond and Yèche, Christophe and Palanque-Delabrouille, Nathalie and Alexander, David M. and Yang, Jinyi and Ahlen, Steven and Bailey, Stephen and Brooks, David and Cai, Zheng and Chabanier, Solène and Davis, Tamara M. and Dawson, Kyle and de la Macorra, Axel and Dey, Arjun and Dey, Biprateep and Eftekharzadeh, Sarah and Eisenstein, Daniel J. and Fanning, Kevin and Font-Ribera, Andreu and Gaztañaga, Enrique and Gontcho, Satya Gontcho A. and Gonzalez-Morales, Alma X. and Guy, Julien and Herrera-Alcantar, Hiram K. and Honscheid, Klaus and Ishak, Mustapha and Jiang, Linhua and Juneau, Stephanie and Kehoe, Robert and Kisner, Theodore and Kovács, Andras and Kremin, Anthony and Lan, Ting-Wen and Landriau, Martin and Guillou, Laurent Le and Levi, Michael E. and Magneville, Christophe and Martini, Paul and Meisner, Aaron M. and Moustakas, John and Muñoz-Gutiérrez, Andrea and Myers, Adam D. and Newman, Jeffrey A. and Nie, Jundan and Percival, Will J. and Poppett, Claire and Prada, Francisco and Raichoor, Anand and Ravoux, Corentin and Ross, Ashley J. and Schlafly, Edward and Schlegel, David and Tan, Ting and Tarlé, Gregory and Zhou, Rongpu and Zhou, Zhimin and Zou, Hu},
  month = aug,
  year = {2022},
}

@article{raichoor_target_2023,
  title = {Target {Selection} and {Validation} of {DESI} {Emission} {Line} {Galaxies}},
  volume = {165},
  issn = {0004-6256, 1538-3881},
  url = {http://arxiv.org/abs/2208.08513},
  doi = {10.3847/1538-3881/acb213},
  language = {en},
  number = {3},
  urldate = {2026-10-28},
  journal = {The Astronomical Journal},
  author = {Raichoor, A. and Moustakas, J. and Newman, Jeffrey A. and Karim, T. and Ahlen, S. and Alam, Shadab and Bailey, S. and Brooks, D. and Dawson, K. and Macorra, A. de la and Mattia, A. de and Dey, A. and Dey, Biprateep and Dhungana, G. and Eftekharzadeh, S. and Eisenstein, D. J. and Fanning, K. and Font-Ribera, A. and Garcia-Bellido, J. and Gaztanaga, E. and Gontcho, S. Gontcho A. and Guy, J. and Honscheid, K. and Ishak, M. and Kehoe, R. and Kisner, T. and Kremin, A. and Lan, Ting-Wen and Landriau, M. and Guillou, L. Le and Levi, Michael E. and Magneville, C. and Martini, P. and Meisner, Aaron M. and Myers, Adam D. and Nie, Jundan and Palanque-Delabrouille, N. and Percival, W. J. and Poppett, C. and Prada, F. and Ross, A. J. and Ruhlmann-Kleider, V. and Sabiu, C. G. and Schlafly, E. F. and Schlegel, D. and Tarle, Gregory and Weaver, B. A. and Yeche, Christophe and Zhou, Rongpu and Zhou, Zhimin and Zou, H.},
  month = mar,
  year = {2023},
  note = {arXiv:2208.08513 [astro-ph.CO]},
  pages = {126},
}

@article{Bordoloi2010,
  author  = {Bordoloi, R. and Lilly, S. J. and Amara, A.},
  title   = {Photo-z performance for precision cosmology},
  journal = {Monthly Notices of the Royal Astronomical Society},
  year    = {2010},
  volume  = {406},
  number  = {2},
  pages   = {881--895},
  doi     = {10.1111/j.1365-2966.2010.16765.x}
}

@article{Bonnett2015,
  author  = {Bonnett, Christopher},
  title   = {Using neural networks to estimate redshift distributions. An application to CFHTLenS},
  journal = {Monthly Notices of the Royal Astronomical Society},
  year    = {2015},
  volume  = {449},
  number  = {1},
  pages   = {1043--1056},
  doi     = {10.1093/mnras/stv230}}




\appendix

\label{sec:appendix}
\section{Supplementary figures}
\label{app:point estimation}

\begin{figure*}
    \centering
    \includegraphics[width=0.45\linewidth]{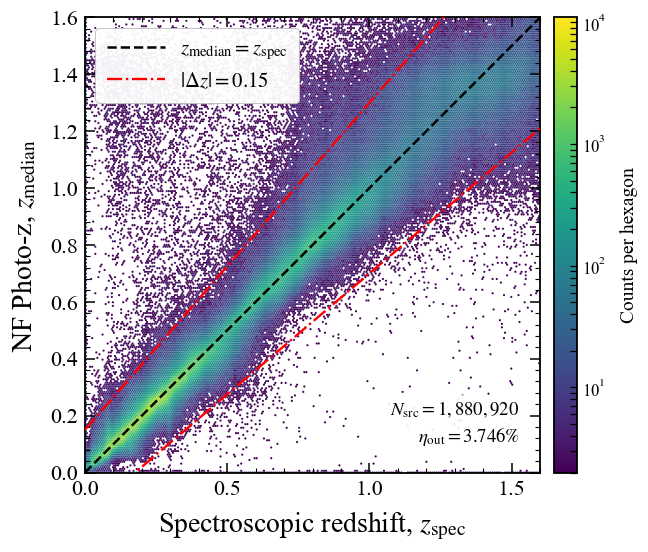}
     \includegraphics[width=0.45\linewidth]{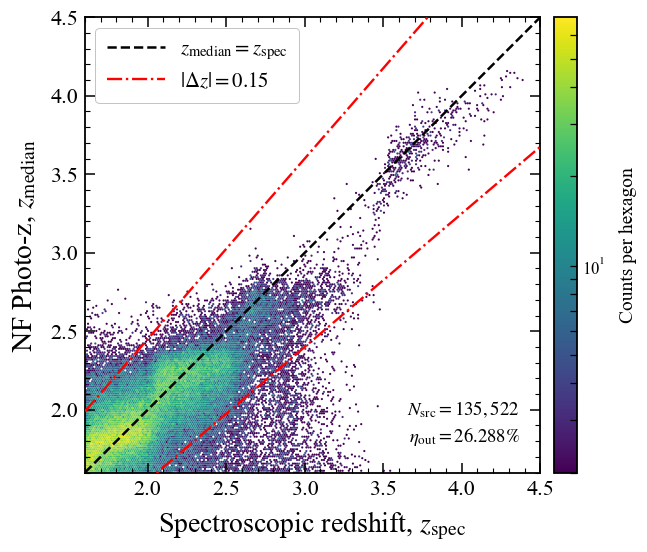}
    \includegraphics[width=0.45\linewidth]{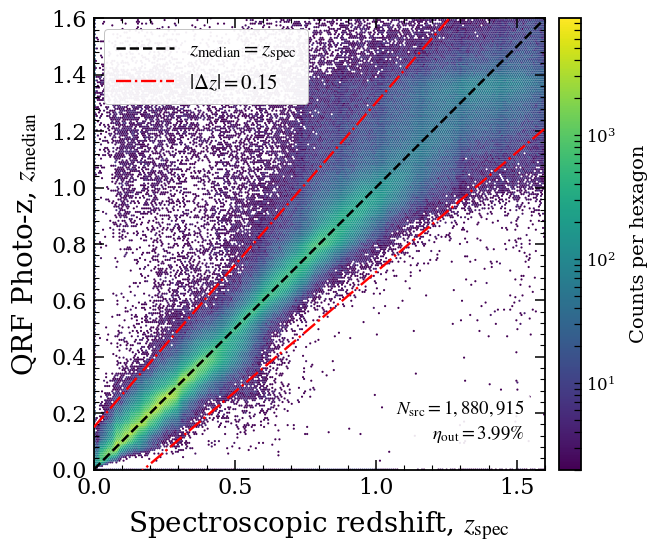}
     \includegraphics[width=0.45\linewidth]{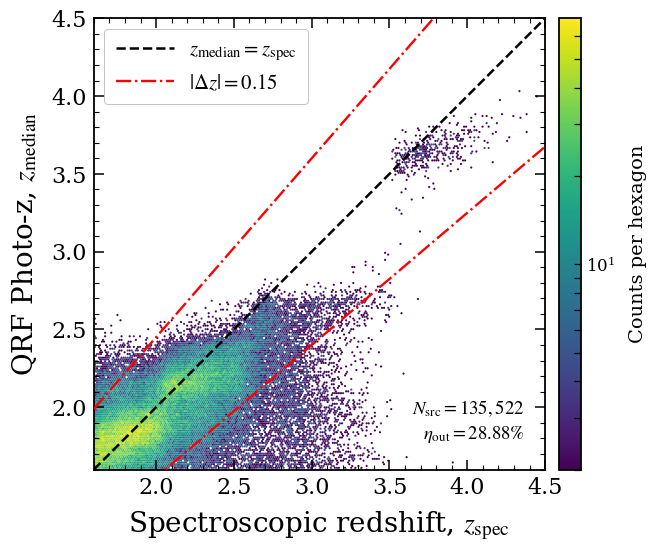}
    \includegraphics[width=0.45\linewidth]{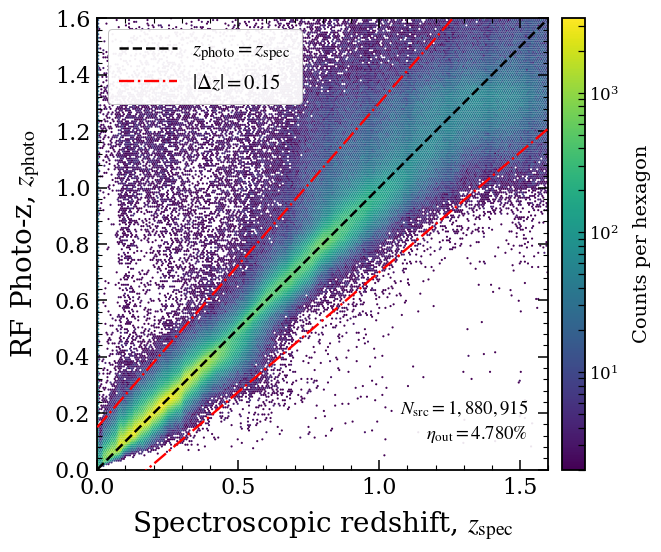}
    \includegraphics[width=0.45\linewidth]{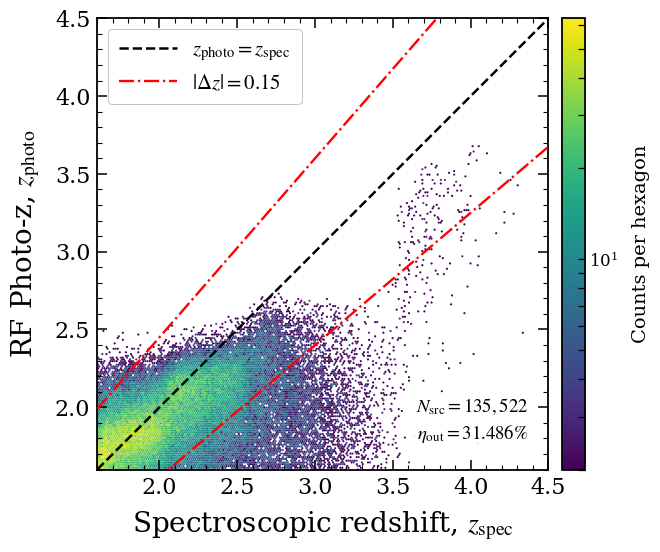}
                \caption{\(z_{\rm spec} -z_{\rm median}\) scatter plots for 1,880,915 sources with \(0.0<z_{\rm spec}<1.6\) and for 135,522 sources with $z_{\rm spec}>1.6$, showing the Normalizing Flow (upper panel), Quantile Random Forest (middle panel), and Random Forest (lower panel). The NF redshift estimate is obtained from the median of each posterior PDF, while the QRF point estimate corresponds to the median derived from 99 uniformly spaced quantile levels between 0.01 and 0.99. The dot dash orange lines indicate the outlier threshold, \(|\Delta z|=0.15\), while the dashed black line shows the one-to-one relation, \(z_{\rm spec}=z_{\rm photo}\). The colour scale indicates the number density of targets. In redshift range $z_{\rm{spec}}>1.6$, the NF predictions are distributed more symmetrically around the one-to-one relation ($z_{\rm phot}=z_{\rm spec}$), while the QRF and RF predictions exhibit more pronounced asymmetric deviations.}
    \label{fig:QRF and RF match}
\end{figure*}

We present the $z_{\rm spec}$--$z_{\rm phot}$ scatter relation for the low redshift $0<z_{\rm{spec}}<1.6$ and high redshift $z_{\rm{spec}}>1.6$  for three models in Fig.~\ref{fig:QRF and RF match}. While the low redshift bin results are visually very similar among the models, the NF results for the high redshift bin are slightly more symmetrical along the diagonal relative to the others.

\begin{figure*}
    \centering
    \includegraphics[width=0.75\linewidth]{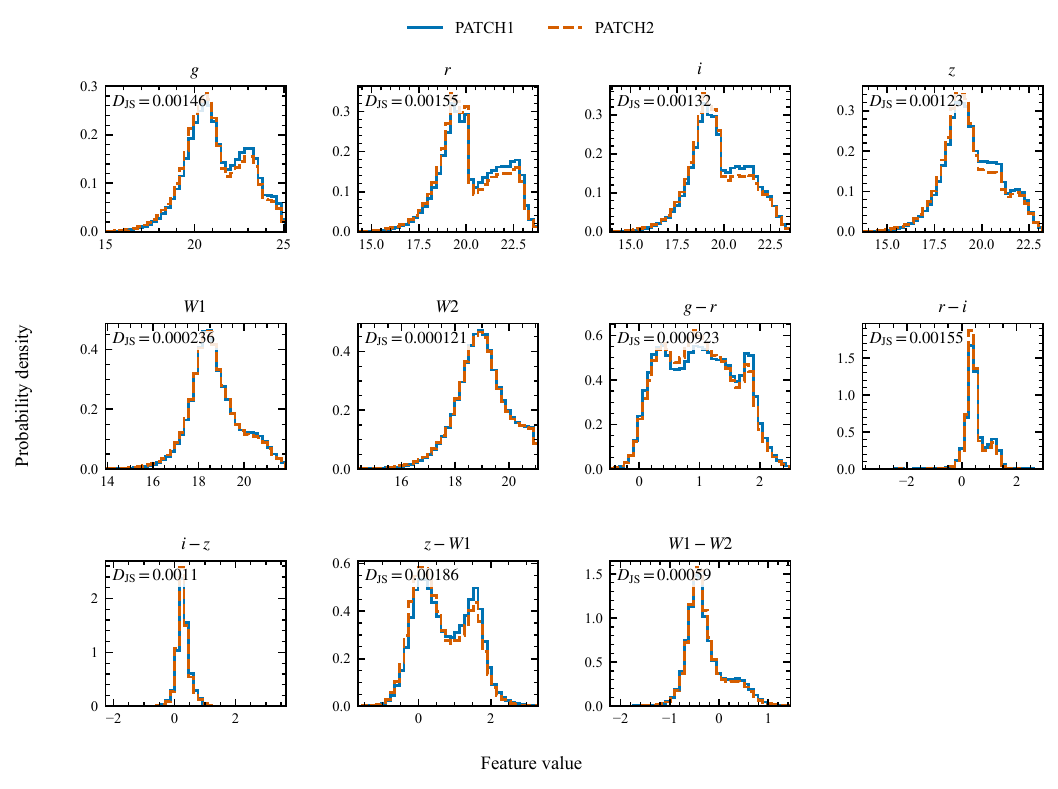}
     \includegraphics[width=0.75\linewidth]{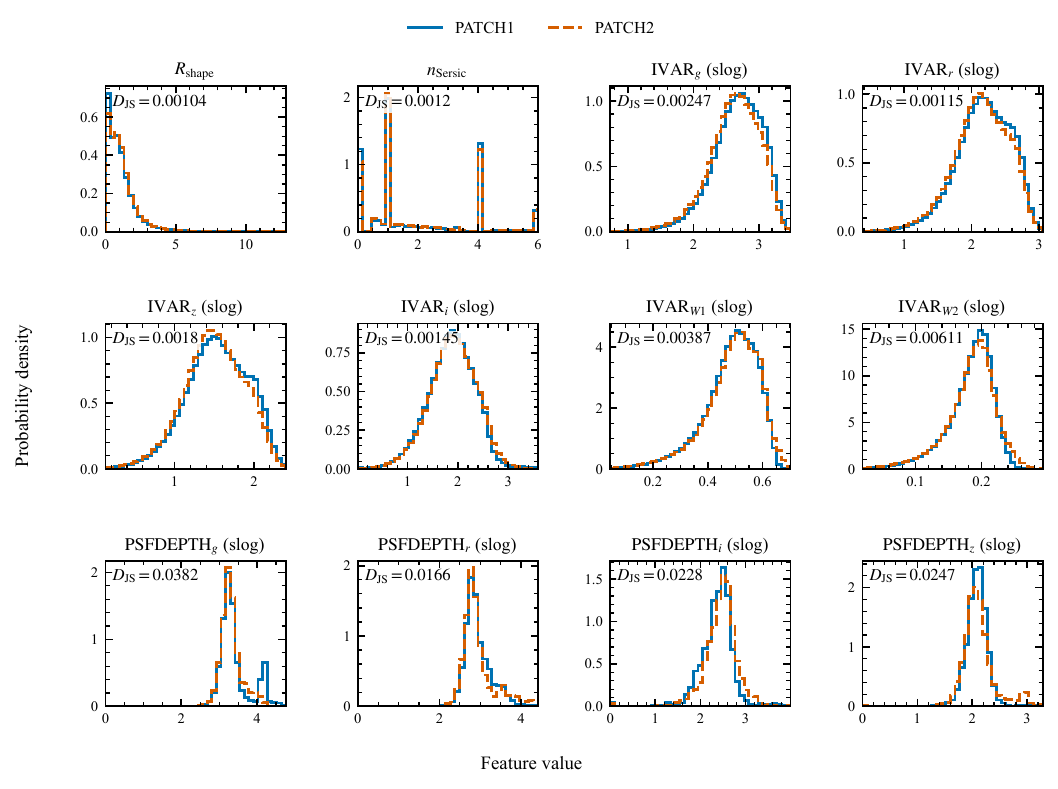}
   \caption{Comparison of the 23 photometric features distributions between training, validation footprint (1,215,914 sources) and testing footprint (2,028,371 sources), including magnitudes ($g$, $r$, $i$, $z$, $W1$, and $W2$), five colour indices ($g-r$, $r-i$, $i-z$, $z-W1$, and $W1-W2$), morphological parameters (SHAPE\_R and SERSIC), flux inverse variances in six bands, and PSF depths in four optical bands. The blue solid and orange dashed lines represent the training, validation patch and testing patch, respectively. The histograms are normalised by the total number of sources in each patch and the corresponding bin widths. For visualization, the displayed range of each feature is determined by the minimum 0.05th and maximum 99.9th percentiles of the two samples. A signed logarithmic transformation, $\operatorname{sgn}(x)\log_{10}(1+|x|)$, is applied to the flux inverse variances and PSF depths to compress their ranges. The Jensen–Shannon divergence, $D_{\rm JS}$, indicated in each panel, is calculated using the full samples with 100 common histogram bins, including boundary bins that retain probability outside the displayed ranges. The two patches exhibit broadly similar distributions, with relatively larger differences in the PSF depths.}
    \label{fig:featuredistributions}
\end{figure*}

Fig.~\ref{fig:featuredistributions} compares the distributions of the input photometric features between training, validation patch and testing patch. The distributions are normalised independently, and the Jensen–Shannon divergence, $D_{\rm JS}$, is used to quantify their differences. The two patches exhibit broadly similar marginal distributions of magnitudes and colours, with relatively larger differences in the PSF depths, particularly in the \(g\) band.


\section{percentile-Based tomographic $N(z)$ test} 

In Section~\ref{sec:tomoN(z)}, we find that the NF model substantially outperforms the QRF model in reconstructing the tomographic redshift distribution for the high-redshift selection $z_{\rm median}>2.5$. However, the two models select substantially different numbers of sources, with approximately 12,000 sources for NF and 7,000 for QRF. To investigate whether this difference in sample size affects the comparison, we perform an additional tomographic test using equal-sized high-redshift samples.

Specifically, we independently select the top 0.5 per cent of sources ranked by their predicted median redshifts, $z_{\rm median}$, for each model. The resulting lower redshift thresholds are $z_{\rm median}=2.55$ for NF and $z_{\rm median}=2.40$ for QRF. This percentile-based selection ensures an identical sample size of 10,142 sources for both models, although the selected samples do not necessarily contain the same objects. For each model, the reconstructed $N(z)$ is compared with the spectroscopic redshift distribution of its corresponding selected sample.

\begin{figure}
    \centering
    \includegraphics[width=0.90\linewidth]{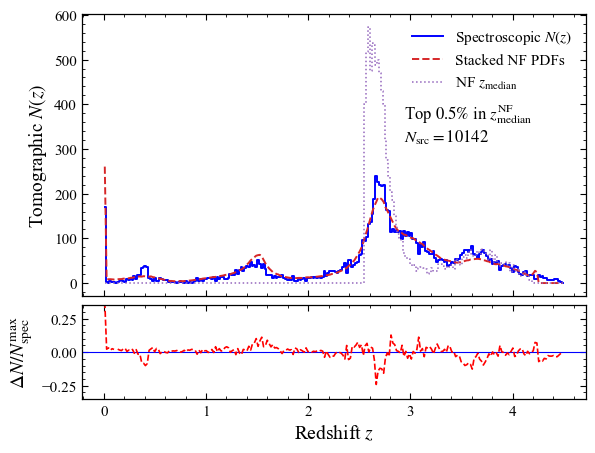}
    \includegraphics[width=0.90\linewidth]{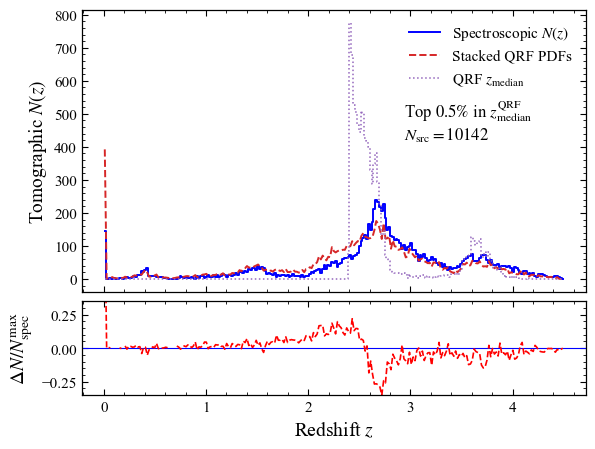}
    \caption{ Equal-sample-size robustness test for high-redshift tomographic $N(z)$ reconstruction using the top 0.5 percent of sources ranked by their predicted median redshifts, $z_{\rm median}$, independently for the NF (upper panel) and QRF (lower panel). The solid blue lines show the spectroscopic reference distributions, the dashed red lines show the reconstructed distributions obtained by stacking the corresponding predicted PDFs, and the dotted purple lines show the distributions of the posterior-median estimates. The lower sub-panels show the residuals, $\Delta N/N_{\rm spec}^{\rm max}$, relative to the corresponding spectroscopic distributions.}
    \label{fig:NzhighQRF}
\end{figure}

The resulting tomographic redshift distributions are presented in Fig.~\ref{fig:NzhighQRF}. The stacked NF PDFs reproduce the spectroscopic $N(z)$ more accurately than the QRF PDFs, particularly around the main high-redshift peak. Quantitatively, NF achieves a distribution overlap of approximately 90$\%$, an absolute mean-redshift bias of $|\Delta z_{\rm mean}|=0.11$, a Wasserstein distance of 0.11, and a Jensen–Shannon (JS) divergence of 0.013. In comparison, QRF yields an overlap of approximately 82$\%$, $|\Delta z_{\rm mean}|=0.24$, a Wasserstein distance of 0.25, and a JS divergence of 0.024. These results are consistent with the findings of the original high-redshift tomographic analysis, demonstrating that the superior $N(z)$ reconstruction achieved by NF persists when the sample sizes are matched, although differences in the selected source populations may also contribute to the observed performance.

\section{Sensitivity of Photometric features in NF model}
\label{app:features}
To investigate the sensitivity of the trained NF model to individual input features, we perform an input-gradient saliency analysis~\citep{simonyan2014deep}. For each source $i$, we define the input feature vector
$\mathbf{x}_i=(x_{i1},\ldots,x_{id})$, where $x_{ij}$
denotes the $j$-th photometric feature of galaxy $i$.
The gradient-based saliency is then calculated as:
\begin{equation}
S_{ij} =
\left|
\frac{\partial \log p_\theta(z_{{\rm spec},i}\mid\mathbf{x}_i)}
{\partial x_{ij}}
\right|.
\label{eq:saliency}
\end{equation}
The gradients are evaluated using the standardised input features, with the trained model parameters held fixed. A larger saliency indicates greater local sensitivity of the predicted log-probability density to variations in the corresponding input feature.

\begin{figure*}
    \centering
     \includegraphics[width=0.95\linewidth]{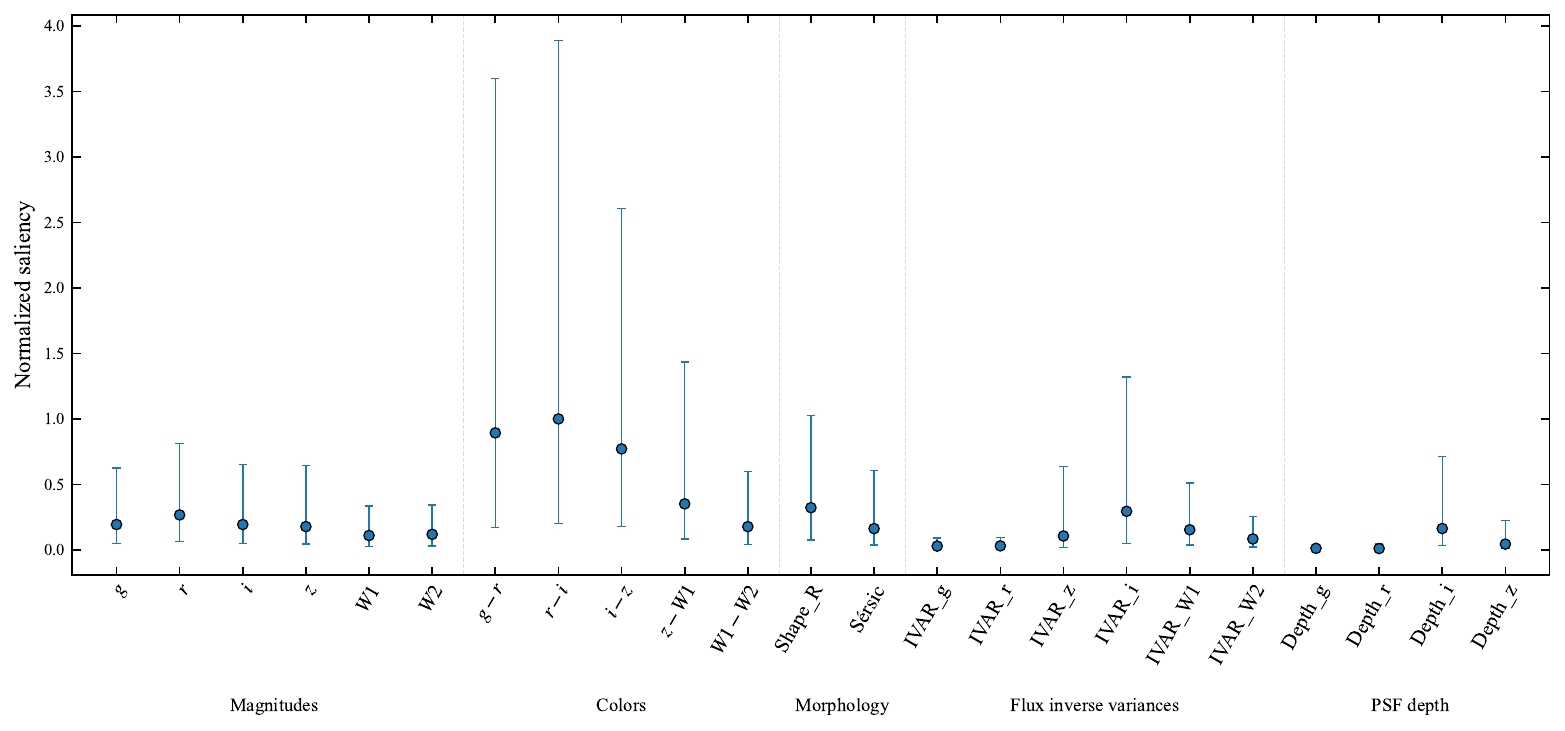}
    \caption{Feature saliency of the trained NF model for the input photometric parameters using 1.1 million training sources. Points show the median absolute input-gradient saliency of each feature, normalised by the maximum median value, while the error bars indicate the 16th–84th percentile range across individual sources. The features are grouped into magnitudes, colours, morphology, flux inverse variances, and PSF depths, with vertical dashed lines separating the groups. A larger saliency indicates greater local sensitivity of the predicted redshift PDF to the corresponding input feature.}
    \label{fig:saliency}
\end{figure*}

For each feature, we compute the median absolute saliency across all evaluated sources, $\widetilde{S}_j = {\rm median}_i(S_{ij})$, and normalise it by the largest median saliency among the input features:
\begin{equation}
S_j^{\rm norm} =
\frac{\widetilde{S}_j}
{\max_k \widetilde{S}_k}.
\end{equation}
The resulting normalised saliencies are presented in Fig.~\ref{fig:saliency}, with error bars indicating the 16th--84th percentile ranges of the individual-source saliencies, normalised by the same factor.

The NF exhibits the strongest sensitivity to colour features, particularly $r-i$, followed by $g-r$ and $i-z$. Individual magnitudes also show appreciable sensitivity, suggesting that the model responds to both colour and magnitude information. The morphological parameters exhibit moderate saliency, whereas the flux inverse variances and PSF depths generally show lower values. The relatively broad 16th--84th percentile ranges of the colour features, especially $r-i$, indicate substantial source-to-source variations in their local influence on the predicted log-probability density.

However, a relatively low gradient saliency does not necessarily imply that a feature has a negligible contribution to the overall photo-$z$ performance. As a complementary analysis, we investigate the impact of individual photometric bands on the point-estimation performance of the RF model. For each band-ablation test, we remove the corresponding magnitude and compute the colour features using the remaining bands. For example, removing the $r$ band replaces $g-r$ and $r-i$ with $g-i$. The resulting performance metrics are summarised in Table~\ref{tab:rf_band_ablation}.

Removing either the $g$ or $r$ band leads to the largest degradation in both $\sigma_{\rm NMAD}$ and the outlier fraction, highlighting the importance of optical photometric information for redshift estimation. Removing the $i$ or $z$ band produces a more moderate degradation. Interestingly, excluding the $W1$ band results in a relatively small increase in $\sigma_{\rm NMAD}$ but a substantial increase in the outlier fraction, comparable to the effects of removing the $g$ or $r$ band. In contrast, removing $W2$ produces smaller changes in both metrics. These results suggest that features with relatively low gradient saliency may still contribute significantly to the overall predictive performance, particularly in reducing catastrophic outliers. Although the NF saliency and RF ablation analyses probe different aspects of feature sensitivity and are performed using different models, they provide complementary insights into the role of photometric information in photo-$z$ estimation.

\begin{table}
\centering
\caption{
Random Forest point-estimation performance for all sources ($\sim2.0$ million targets in $0.0<z_{\rm{spec}}<4.5$) after removing individual photometric bands from the input feature set. For each ablation, the corresponding magnitude and all colours containing that band are removed, and the colour set is reconstructed from the remaining adjacent bands (e.g. removing $r$ replaces $g-r$ and $r-i$ with $g-i$). The full RF model is shown for reference.
}
\begin{tabular}{lcccc}
\toprule
Removed band &
 bias($\times10^{-4}$) &
$\sigma_{\rm NMAD}(10^{-2})$ &
Outlier fraction (\%) \\
\midrule
None &4.65 &2.92 &6.58 \\
$g$ &-4.77 &3.56 &7.46 \\
$r$ &-0.33 &3.51 &7.47 \\
$i$ &7.79 &3.24&6.91 \\
$z$ &6.88 &3.26 &7.06 \\
$W1$ &6.84 &3.09 &7.41 \\
$W2$ &5.82 &3.07 &7.10 \\
\bottomrule
\end{tabular}
\label{tab:rf_band_ablation}
\end{table}

\clearpage

\bsp	
\label{lastpage}
\end{document}